\documentclass[conference]{IEEEtran}

\usepackage[export]{adjust box}
\usepackage{pgf}
\usepackage{booktabs}
\usepackage{balance}
\usepackage{paralist}
\usepackage{amsmath}

\usepackage[colorinlistoftodos, textwidth=2cm,textsize=tiny]{todonotes}

\usepackage{url}
\usepackage{graphicx}
\usepackage{float}
\usepackage{array}
\usepackage{multirow}
\usepackage{setspace}
\usepackage{textcomp}

\usepackage{float}
\usepackage[caption = false]{subfig}

\usepackage{dblfloatfix}

\usepackage{xcolor,colortbl}

\definecolor{Gray}{gray}{0.85}
\newcolumntype{a}{>{\columncolor{Gray}}c}
\newcolumntype{C}[1]{>{\centering\arraybackslash}m{#1}}

\newcommand{\spara}[1]{\smallskip\noindent{\bf #1}}

\usepackage{fancyhdr} 
\usepackage{kantlipsum} 
\fancypagestyle{plain}{ 
\fancyhf{} 
\fancyhead[C]{Conference on \LaTeX} 
 
} 
\usepackage{eso-pic} 

\begin{document}

\AddToShipoutPictureBG*{ 
\AtPageUpperLeft{ 
\setlength\unitlength{1in} 
\hspace*{\dimexpr0.5\paperwidth\relax}
\makebox(0,-0.75)[c]{\textbf{2018 IEEE/WIC/ACM International Conference on Web Intelligence (WI'18)}}}} 

\title
{Ranking of Social Media Alerts with \\ Workload Bounds in Emergency Operation Centers} 

\author{\IEEEauthorblockN{Hemant Purohit}
\IEEEauthorblockA{\textit{George Mason University}\\
Fairfax, VA, USA \\
hpurohit@gmu.edu}
\and
\IEEEauthorblockN{Carlos Castillo}
\IEEEauthorblockA{\textit{Universitat Pompeu Fabra}\\
Barcelona, Spain \\
chato@acm.org}
\and
\IEEEauthorblockN{Muhammad Imran}
\IEEEauthorblockA{\textit{Qatar Computing Research Institute}\\
Doha, Qatar \\
mimran@hbku.edu.qa}
\and 
\IEEEauthorblockN{Rahul Pandey}
\IEEEauthorblockA{\textit{George Mason University}\\
Fairfax, VA, USA \\
rpandey4@gmu.edu}
}


\maketitle



\begin{abstract}
Extensive research on social media usage during emergencies has shown its value to provide 
life-saving information, if a mechanism is in place to filter and prioritize messages. 
Existing ranking systems can provide a baseline for selecting which updates or alerts to push to emergency responders.  
However, prior research has not investigated in depth how many and how often should these updates be generated, considering a given bound on the workload for a user due to the limited budget of attention in this stressful work environment.  

This paper presents a novel problem and a model to quantify the relationship between the performance metrics of ranking systems (e.g., recall, NDCG) and the bounds on the user workload. We then synthesize an alert-based ranking system that enforces these bounds to avoid overwhelming end-users. We propose a Pareto optimal algorithm for ranking selection that adaptively determines the preference of top-\textit{k} ranking and user workload over time.     
We demonstrate the applicability of this approach for Emergency Operation Centers (EOCs) by performing an evaluation based on real world data from six crisis events. We analyze the trade-off between recall and workload recommendation across periodic and realtime settings.  
Our experiments demonstrate that the proposed ranking selection approach can improve the efficiency of monitoring social media requests while optimizing the need for user attention.  
%
\end{abstract} 



\section{Introduction}
\label{sec:intro}

Social media analytics has become a mainstream part of organizational workflows and services in all kinds of organizations, including governments and for-profits.  
The use of social media in organizations has demonstrated improvements in their customer relations and services. 
Likewise, for emergency management, a substantive body of research has shown how response agencies and nonprofits can monitor social media for situational awareness~\cite{arc_2012_survey,hughes2012evolving}. 

However, due to the characteristics of ``big crisis data''~\cite{castillo2016big}, which includes high volume and velocity, there are many challenges in monitoring social media message streams.   
These messages have varied degrees of information and noise that may not be of potential value for the response, ranging from actionable requests or offers of help~\cite{purohit2013emergency,he2017signal} and unsubstantiated rumors~\cite{starbird2014rumors} to gratitude and advertisement~\cite{purohit2018socialeoc,imran2015processing}. Thus, finding the relevant social media updates is a critical concern for emergency management services.   
\begin{table}
\centering
\caption{Motivation to study a relationship between ranking metrics and workload. The design expectation of high recall of ranked alerts with a low workload of monitoring implies the lesser need of attention from the time-crunched end users.} 
\begin{tabular}{ c c | c c }
	& & \multicolumn{2}{c}{\textbf{\underline{Workload}}}   \\
               &	   & \textit{High} & \textit{Low} \\ 
     			 	\hline  
\multirow{2}{4em}{\textbf{\underline{Recall}}} & \textit{High} & 		inefficient & desired \\  
           	 & \textit{Low}  & worst & ineffective        
\end{tabular}
\label{tab:motivation}
\vskip -0.22in
\end{table}

Existing alert-based systems~\cite{avvenuti2014ears,aslam2015trec} can provide a solution to generate an alert every time a relevant (sub-) event is detected through social media. However, given a highly dynamic situation, too many alerts can be triggered during a short time window. Thus, such alerts intended to help in social media monitoring would rather serve as distractions to the multitasking EOC personnel and hamper their work during 
a time-critical event~\cite{hiltz2014use,plotnick2015red}. 
%
 Information Retrieval literature 
 provides different types of ranking systems~\cite{baeza2011ribeiro}, which could be employed to select the multiple relevant items for various emergency services. 
For instance, periodic top-\textit{k} retrieval systems can be used for batch-based processing of social media, while continuous systems with push-based and pull-based processing can be used for summarizing trends and sending alerts~\cite{aslam2015trec,avvenuti2014ears,rudin2009p}.  
In particular, push-based alert systems are relevant in the context of an emergency operation center (EOC). However, it is unclear if a ``one-size-fits-all'' solution of any alert system would be applicable,    
 given the highly stressful work environment of emergency workers that leads to a limited budget of user attention/workload. Table~\ref{tab:motivation} shows the trade-off and the design expectation for the relationship between the system and user workload metrics. An adaptive alert system with high recall and low required workload would be efficient and effective to minimize the waste of user efforts. 
\vskip 0.1in
\noindent \textbf{Contributions}. We formulate a novel problem of how to create an alert ranking system that is adaptive to the bounds on user performance, for deciding how many alerts to generate and when. 
We present a novel model of the human-machine interaction to quantify the relationship between the performance metrics of an alert ranking system (e.g., recall) and a user (e.g., workload). 
We also evaluate, using real-world crisis datasets, an alert ranking system that enforces a maximum user workload to avoid overwhelming users and adaptively select an appropriate set of ranked messages. 
%
%

The rest of this paper 
first describes 
related work in Section~\ref{sec:related}. Section~\ref{sec:approach} presents our approach to model the ranking-workload relationship, and an optimization method for selecting a ranking policy. Finally, in Section~\ref{sec:experiments} we demonstrate the applicability of this model by analyzing datasets from 6 crises, before discussing the limitations in Section~\ref{sec:discuss} and our conclusions in Section~\ref{sec:conclusion}. 

\section{Related Work}
\label{sec:related}
Literature on social media in emergency management is vast, for a survey, see~\cite{imran2015processing,castillo2016big}. 
In this section we focus on works that are closely related to our problem of selecting which ranking system for emergency services is appropriate for operational efficiency. 

\subsection{Social media during emergencies}

Improving social media-based emergency response is challenging due to ``big data'' characteristics of high volume, variety, and velocity of social media streams, which overwhelm response services in processing data for relevant information~\cite{castillo2016big}. 
Crisis informatics \cite{palen2016crisis} research has investigated the use of social media during disasters for response services. The prior research identifies a key challenge and a barrier for the efficient use of social media communication channel for response organizations as the information overload on responders~\cite{hiltz2014use,castillo2016big}. Information overload originates from a variety of factors including the large scale of unstructured and noisy nature of social media streams and the lack of time to cognitively process relevant content in social media to prioritize for response.     
In the emergency management domain, Public Information Officers (PIOs) play the crucial role to collect relevant information from public sources for the response agencies or an EOC, by leveraging various information communication technologies including social media~\cite{hughes2012evolving}. 
 In this context, PIOs of formal emergency management services have started using social media channels to communicate effectively with public and source relevant information for actionable intelligence. The recent reports and surveys of formal response organizations \cite{dhs2014using,reuter2017towards} recognize social media as a novel information channel for improving operational response coordination. However, a research question of how and when to effectively monitor social media for finding relevant information remains open. 

\subsection{Push and Pull Ranking Systems}  
We briefly summarize various 
ranking methods for the problem of alert generation in different domains. 

Researchers have investigated diverse techniques for push-based alert systems within the context of disaster management, in particular, using event detection approaches. 
Sakaki et al.~\cite{sakaki2013tweet} proposed a method for near-realtime earthquake event detection and alert dissemination via Twitter. They devised a classification algorithm to monitor tweets for detecting a target event and send out corresponding alert. 
Earle et al.~\cite{earle2012twitter} also evaluated an earthquake detection and alert dissemination procedure solely relying on temporal pattern analysis for the keyword-based tweet-frequency time series.  
Avvenuti et al.~\cite{avvenuti2014ears} developed a burst detection algorithm to promptly identify outbreaking seismic events and automatically 
broadcasted alerts via a dedicated Twitter account and by email notification systems. 
Robinson et al.~\cite{robinson2013sensitive} and Yin et al.~\cite{yin2012using} developed Emergency Situation Awareness platform for earthquake detection in Australia and New Zealand regions using Twitter, which sends email notifications for evidence of earthquakes. 
%
%
Researchers have also designed news feed systems that give top-$k$ alerts, which are relevant for users who subscribe to specific information sources. For instance, Bao and Mokbel~\cite{bao2013georank} developed a location-aware system for news feed ranking, where top-$k$ news feeds were selected based on spatial-temporal proximity and the user preference characteristics.  
The key limitation of the existing alert generation methods within our problem context is that it is not clear when to generate alerts and how many to generate, for efficiently assisting and not obstructing an EOC expert's task. 

\subsection{Summarization Update Systems}  
Another major category of work related to our problem is in the area of data stream summarization. Several researchers have devised methods to generate summarization updates for dynamic top-$k$ relevant items. 
Aslam et al.~\cite{aslam2015trec} defined an information access problem in the context of streaming data and proposed a track in the well-known TREC Challenge. 
The challenge was to develop systems for efficiently monitoring the information associated with an event and broadcast short, relevant, and reliable sentence-length updates about the developing event. 
Kedzie et al.~\cite{kedzie2015predicting} presented a system for update summarization that predicts the salience of
sentences with respect to an event using disaster-specific features including geo-locations and language models, and then bias a clustering algorithm for sentence selection for updates. 
McCreadie et al.~\cite{mccreadie2014incremental} developed  a novel incremental update summarization approach that adaptively alters the volume of content issued as updates over time with respect to the prevalence and novelty of discussions about the event. 
Rudra et al.~\cite{rudra2015extracting} proposed a framework that first classifies tweets to extract situational information and then,  summarizes the information for a user. Their approach factored in the disaster-specific tweet characteristics that contain both situational and non-situational information.  
Nenkova et al.~\cite{nenkova2011automatic} provide an extensive survey on automated summarization methods.  
Our goal is to not develop update-summarization algorithm, but the selection policy for adapting the appropriate behavior of the ranking algorithm for updates as per the end user's workload. 

\vskip 0.1in
\noindent \textit{\textbf{Gap summary}}: While the above-discussed works on alert generation and stream/update summarization methods are related, they do not account for and study the relationship between the number of (top) `k' alerts/updates to generate and the bounds on the user's workload. Thus, none of the existing systems can be directly used to address our problem. 
Instead, these works 
 motivate the novel problem to design a generalizable ranking selection method that is aware of the user workload bounds. 

%

\section{Approach: Workload-Bounded Alert Ranking}    
\label{sec:approach} 
This section first describes our problem formally and then, the solution 
to model the ranking-workload relationship as well as 
select a ranking policy for generating alerts. 

\spara{Problem Statement}. Let 
 $t_{ij}$ be a finite time period from timestamp $t_i$ to $t_j$  ($0 \leq i < j \leq n$),  
 $x_{ij}$ be a finite set of messages generated in $t_{ij}$, 
 $w(k,t_{ij})$ be the required user workload to monitor $k$ messages ($0 < k < |x_{ij}|$) in $t_{ij}$, and 
  $B$ as the bound on maximum user workload in the total time period $[t_0, t_n]$, i.e. $\sum_{ij} w(k,t_{ij}) \leq B$.   
Select 
 a ranking function $R(x_{ij})$ to retrieve top-$k$ items in $t_{ij}$ for alerts   
 such that 
 the ranking-performance metric $M(R(x_{ij}))$ is maximum and 
 the required user workload $w(k,t_{ij})$ is minimum. 

\begin{table}
\centering
\small
\caption{Example of a $RW$ matrix (rows as $k$, columns as $t_{ij}$, and cells as the attainable $R$ecall and $W$orkload tuples).} 
\begin{tabular}{|a|c|c|c|c|c|c|} 
    \hline
    \rowcolor{Gray}
    $k$ &    $t_{ij}$=10 & $t_{ij}$=20 & $t_{ij}$=30 & $t_{ij}$=40 & $t_{ij}$=50 & $t_{ij}$=60  \\
    \hline
    1  & (48, 6) & (34,3)  & (27,2)  & (23,1.5) & (20,1.2) & (18,1)  \\
    2  & (70,12) & (53,6)  & (44,4)  & (38,3)   & (34,2.4) & (30,2)  \\
    .. & ..      & ..      & ..      & ..       & ..       & ..      \\
    10 & (99,60) & (96,30) & (90,20) & (84,15)  & (78,12)  & (73,10) \\ 
    \hline
\end{tabular}
\label{tab:matrix}
\vskip -0.1in
\end{table}

\spara{Solution}. Given the varied types of tasks in EOCs, an alert could be generated for serving different user roles. For a concrete demonstration of our proposed solution to the above problem, we consider the alerts targeted for public communication personnel when a citizen requests to help for a resource or seek information during a disaster. Our solution approach involves three specific steps as described next:  
i.) relevant message identification and ranking, 
ii.) ranking-workload ($RW$) matrix generation, and
iii.) optimal $RW$ policy selection.

\begin{table*}[t!]
  \caption{Summary of datasets for tweets contained in the directed conversations to response agencies on social media, which may contain a potential alert. } 
  \label{tab:dataset} 
  \centering
  \begin{tabular}{p{7cm}|rrr} 
    \hline 
    Event (start-end month/day) &
    \begin{minipage}{2cm}  Tweets \end{minipage} & 
    \begin{minipage}{1.8cm}  Relevant  \end{minipage} &
    \begin{minipage}{1.8cm}  \mbox{Non-Relevant} \end{minipage} \\ 
    \hline 
    Hurricane Sandy 2012 (10/27-11/07)  &  1,153    & 40\% & 60\% \\ 
    Oklahoma Tornado 2013 (05/20-05/29) &  1,513   & 48\% &  52\% \\
    Alberta Floods 2013 (06/16-06/16)   &  2,727        & 28\% &  72\% \\
    Nepal Earthquake 2015 (04/15-05/15) &  2,222    & 18\% &  82\% \\
    Louisiana Floods 2016 (10/11-10/31) &  1,369      & 34\% &  66\% \\ 
    Hurricane Harvey 2017 (08/29-09/15) &  12,742  & 20\% &  80\% \\
  \hline 
\end{tabular}
\vskip -0.1in
\end{table*}

\begin{figure*}
 \centering
 \subfloat[]{\includegraphics[width = 2.3in]{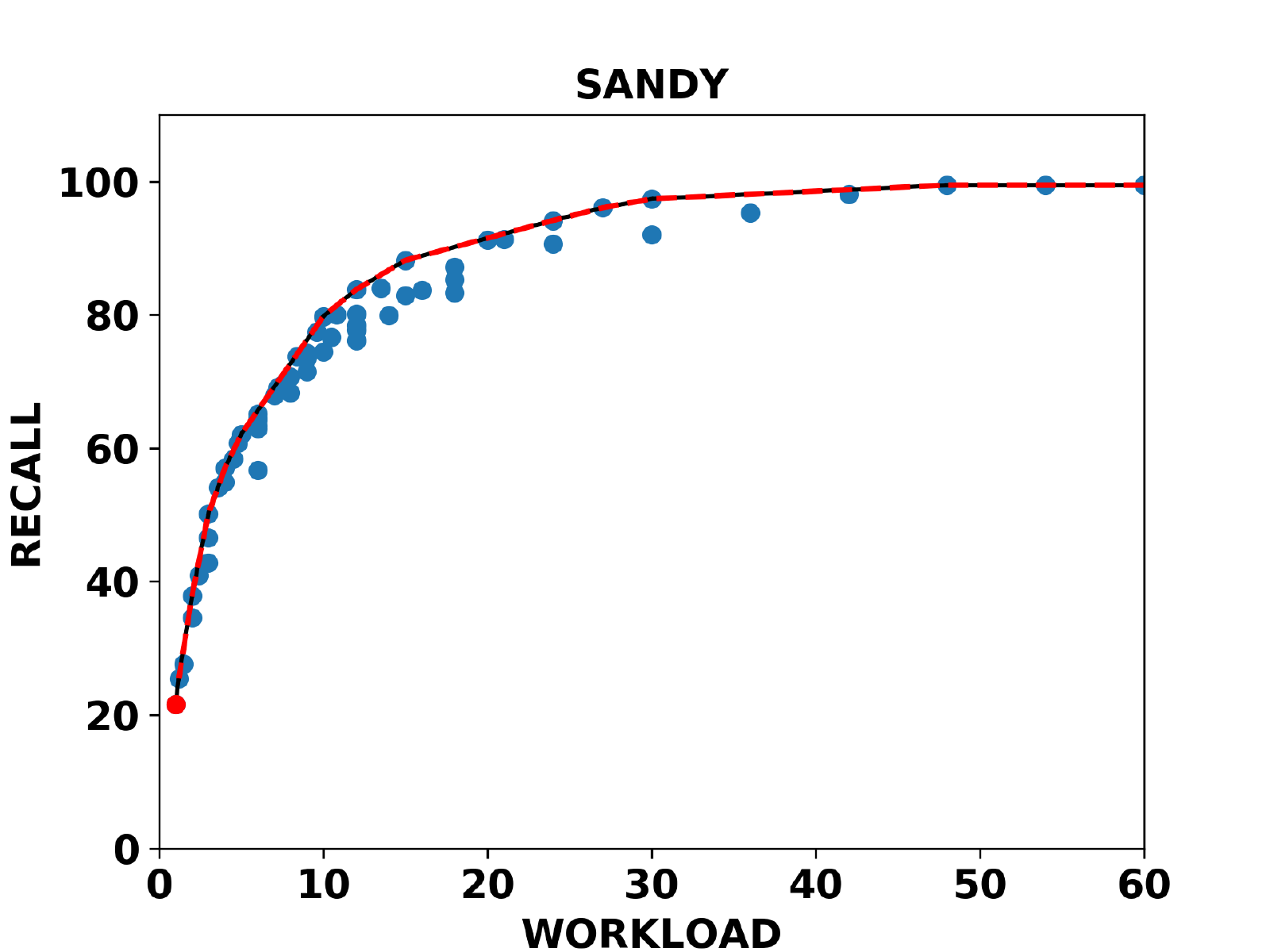}}  
 \subfloat[]{\includegraphics[width = 2.3in]{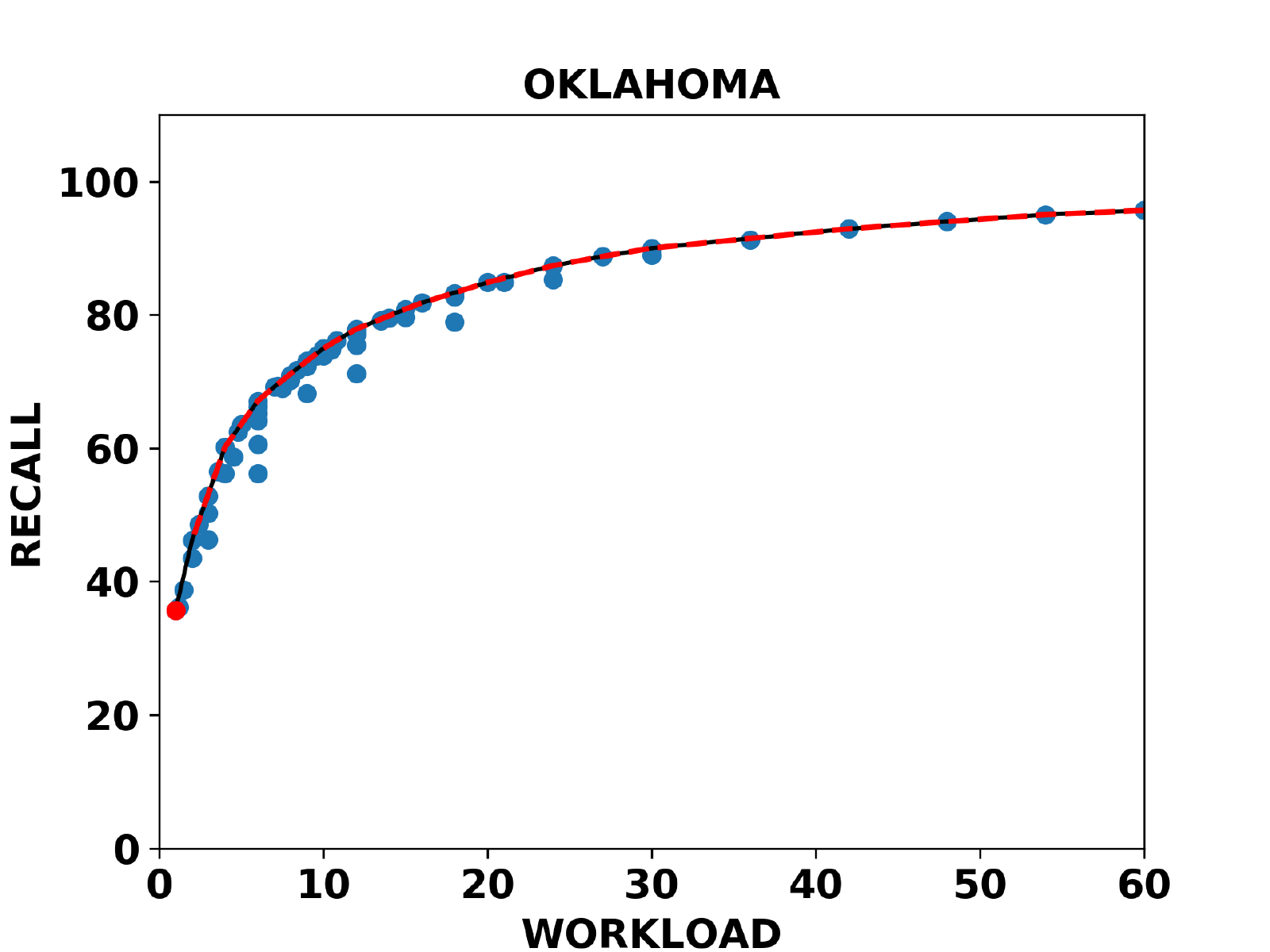}} 
\subfloat[]{\includegraphics[width = 2.3in]{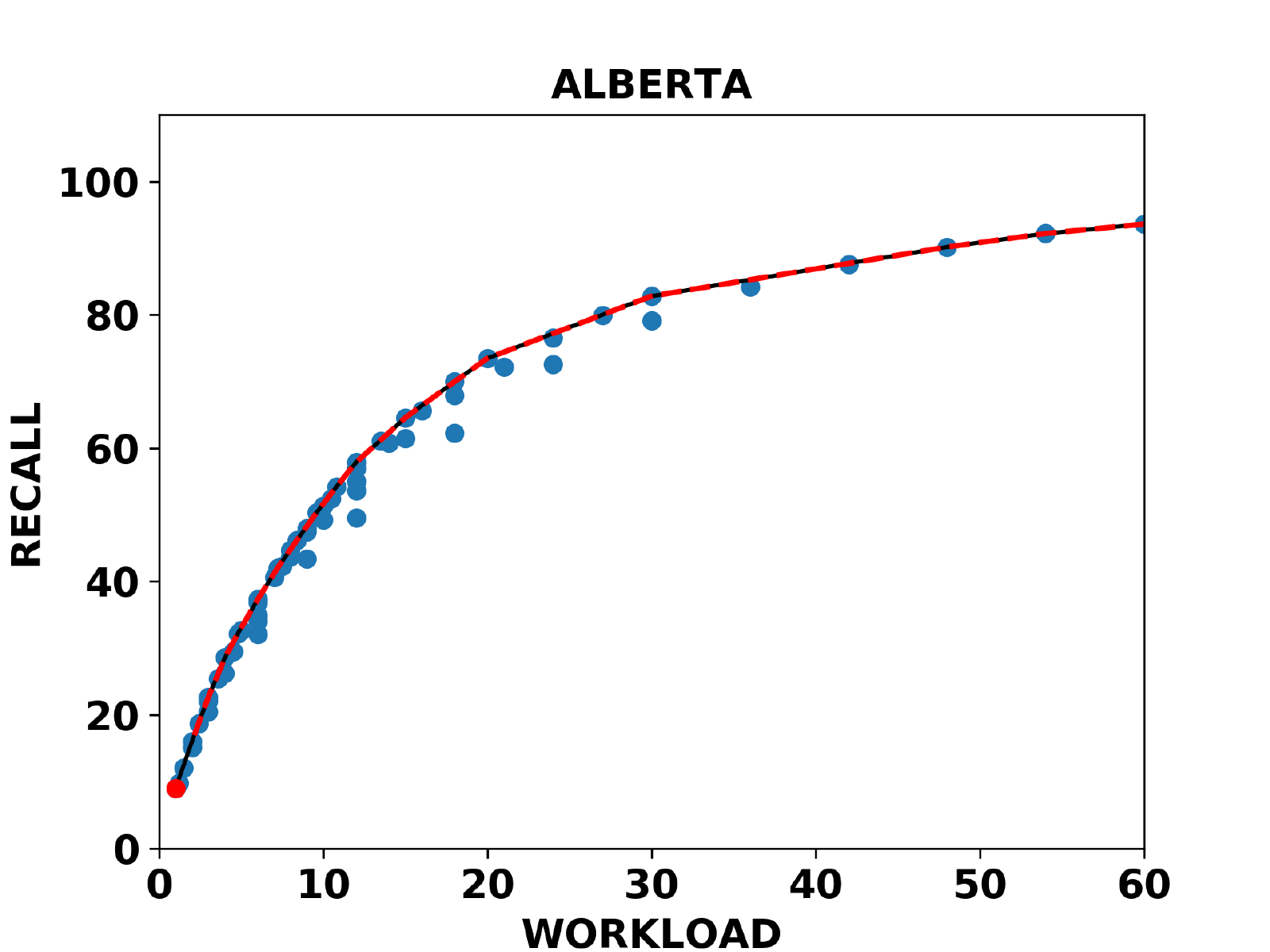}} \\ 
 \subfloat[]{\includegraphics[width = 2.3in]{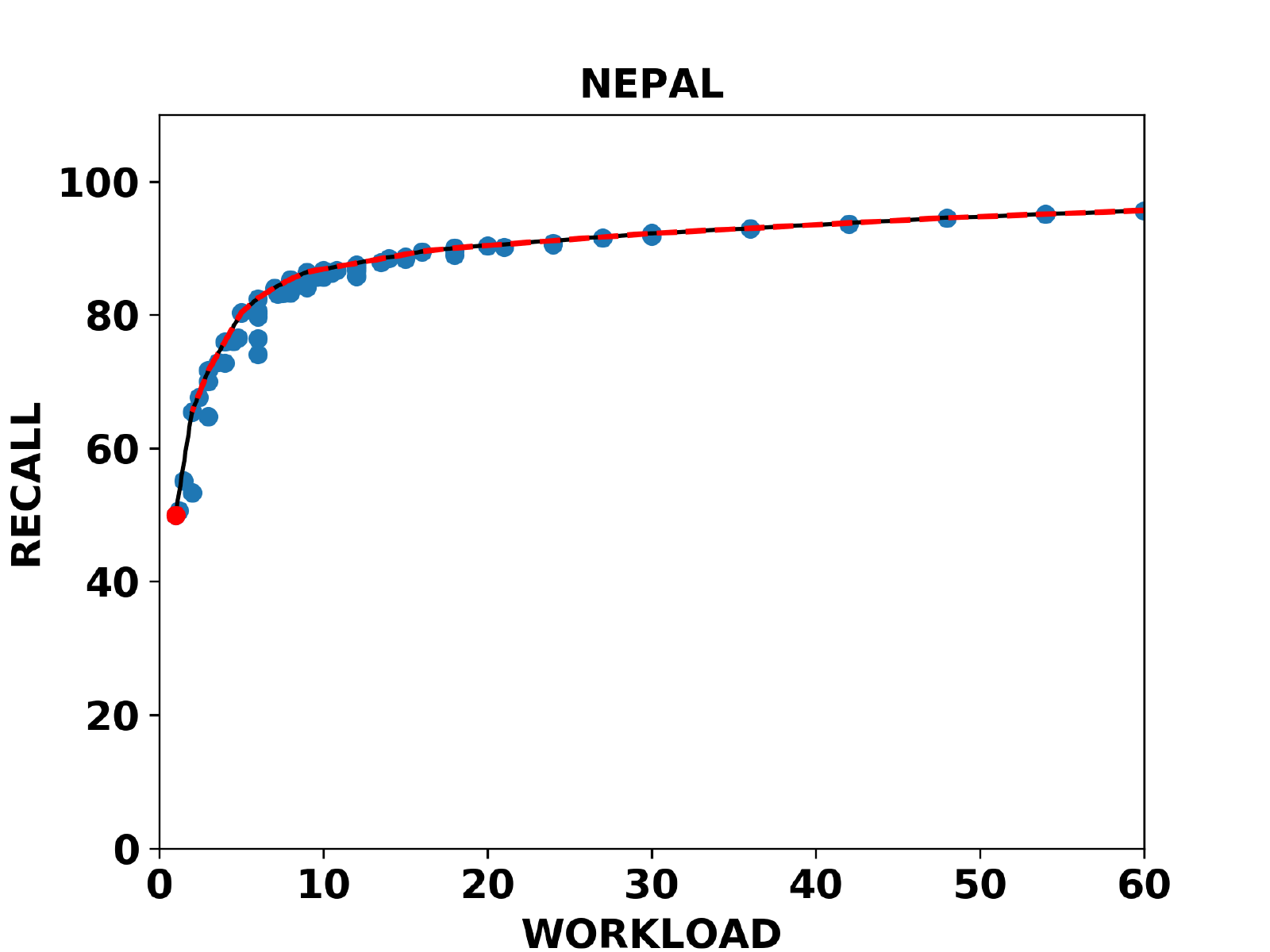}} 
  \subfloat[]{\includegraphics[width = 2.3in]{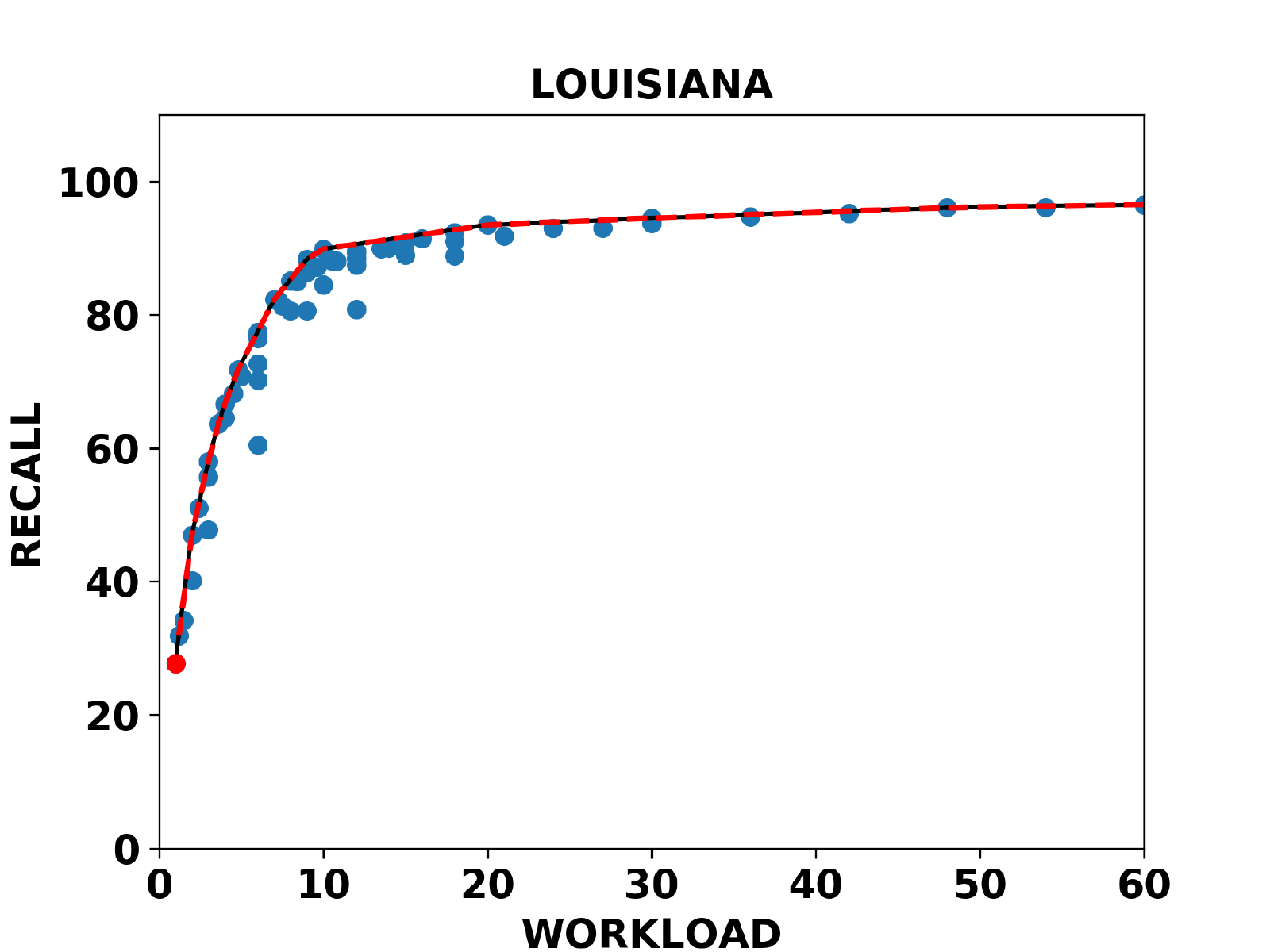}}
 \subfloat[]{\includegraphics[width = 2.3in]{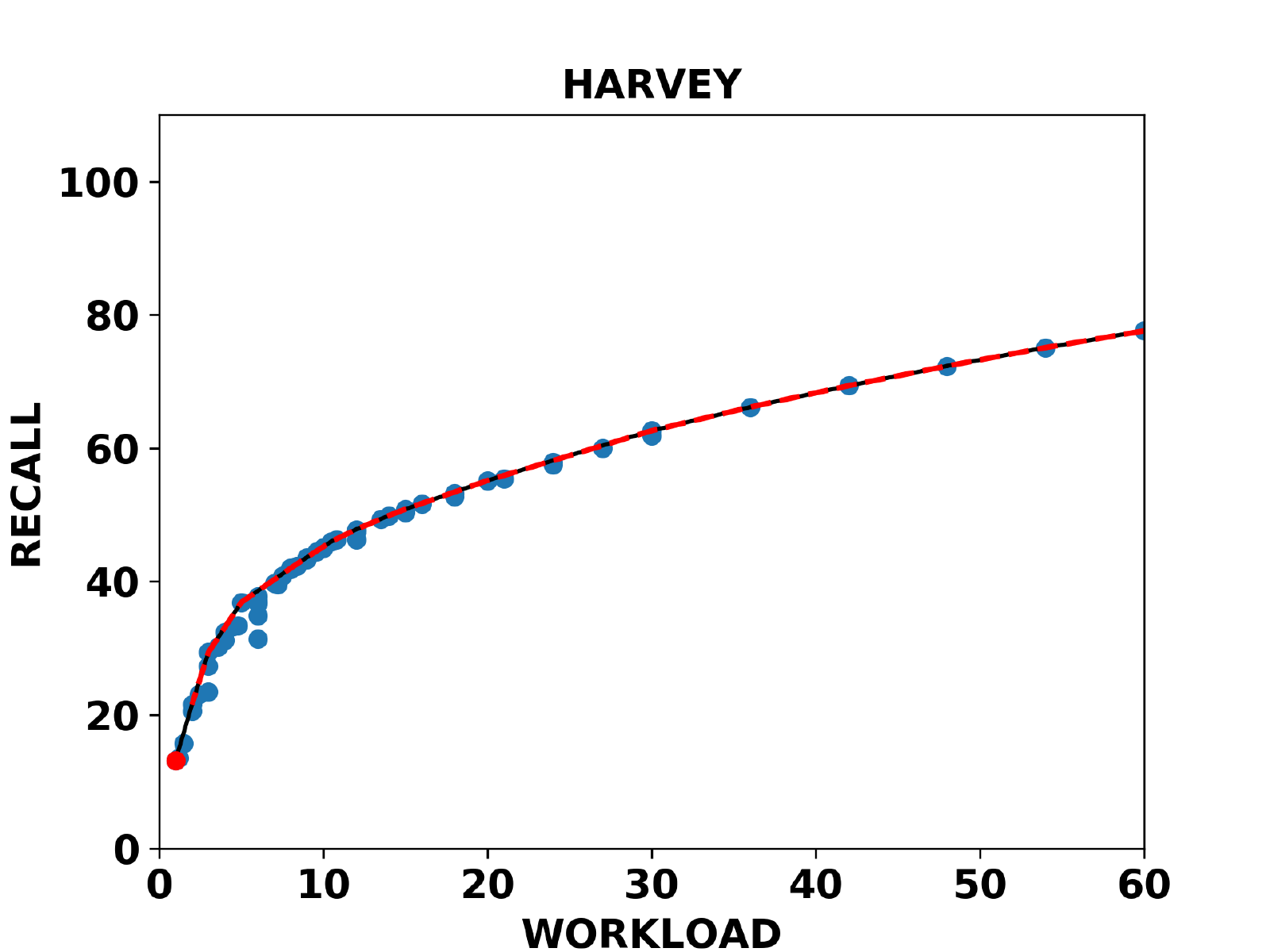}} 
 \caption{Diminishing returns of a workload/attention budget. Red lines: maximum recall achievable given a budget. Blue dots: multiple average recall values for a given budget. \textit{[\textbf{All higher resolution figures available at:} \textcolor{blue}{\protect\url{http://ist.gmu.edu/~hpurohit/informatics-lab/blogs/wi18-workload-aware-ranking.html}}]}} 
\label{fig:workload-chull}
\vskip -0.1in
\end{figure*}

\subsection{Relevant Message Identification \& Ranking}

We consider a general class of emergency service requests as relevant messages for alerts that include \textit{actions}, such as a request for resources (e.g., emergency medical assistance for an injured person) as well as \textit{information} (e.g., a request for a phone number for information on missing people)~\cite{sachdeva2017call,purohit2018socialeoc}. 
We have considered serviceability of messages as the relevance criterion~\cite{purohit2018socialeoc}. The key characteristic of serviceability of a request message for an alert is that it requests a resource that can be provided, or asks a question that can be answered by the service personnel.  
Our approach requires a relevancy classification and ranking for the messages. Thus, we adapted the learning-to-rank methodology~\cite{liu2009learning} and designed a SVM-Rank classifier, using the labeled messages with binary relevance classes provided by the emergency domain experts in the prior research~\cite{purohit2018socialeoc}. For features, we first used Bag-of-Words features that achieved accuracy of only 65\%. Therefore, we resolved to an improved approach for the relevancy classification with better accuracy from our prior work~\cite{purohit2018socialeoc} that used additional features of informative details, such as time, place, or context in the message content. 
Using the relevancy classification and ranking prediction for messages, we compute the ranking metrics for a given set of messages $x_{ij}$ in a period $t_{ij}$ for different types of ranking from top-$1$ to top-$k$ alerts.  

\subsection{Ranking-Workload ($RW$) Matrix Generation}  

We propose a matrix-based model to formalize the relationship between ranking performance metrics and the end user workload. We define a $RW$ matrix, as shown in table~\ref{tab:matrix}, where rows represent the number of top-$k$ alerts to generate and the columns represent the period $t_{ij}$ for the frequency of generating the top-$k$ alerts. The matrix contains 2-tuple values of functions corresponding to the ranking metric and user workload as follows:
\begin{equation}
 		RW(k,t_{ij}) = \; \langle \, M(R(x_{ij})), w(t_{ij},k) \, \rangle
\end{equation} 
where, 
\begin{itemize}
\item  $M(R(x_{ij}))$ is the ranking metric function that computes the performance score for a chosen top-$k$ alert ranking $R(x_{ij})$ of message set $x_{ij}$ in $t_{ij}$, such as Precision@$k$, NDCG@$k$, and Recall@$k$~\cite{baeza2011ribeiro}.  
\item  $w(t_{ij},k)$ is the user workload function that characterizes the notion of the amount of hourly work in industry. We define the user workload as the number of alerts to monitor in $h$ hours (or $h*60$ minutes):     
\begin{equation} 
       w(t_{ij},k) = \; k * ( h / t_{ij} )  \; | \;   1 \leq t_{ij} \leq h        
\end{equation} 
\end{itemize}

 For simplicity, we consider $h=1\; hour$, i.e., 60 minutes and $t_{ij} \in \{10,20,30,40,50,60\} \; minutes$. For instance, $t_{ij}=10$ and $k=5$ imply that an end user will need to monitor top-5 alerts every 10 minutes and the required workload for him will be the cognitive processing of 30 messages per hour. 
 
We consider the top-$k$ alert systems for $k \in [1,2,..,10]$. We constructed the $RW$ matrix for $t_{ij}$ using the ranking metric function as Recall@$k$ for the top-$k$ results from the predicted relevant messages in $x_{ij}$.  

\subsection{Optimal $RW$ Policy Selection for Recommendation} 

Given the multiple choices of workload and desired recall values in the $RW$ matrix, as illustrated in table~\ref{tab:matrix}, it is challenging to determine which combination of the top-$k$ ranking and $t_{ij}$ period be recommended. For instance, (row $k$=1, column $t_{ij}$=60) shows the minimum workload setting (18,1), although with low recall, while (row $k$=10, column  $t_{ij}$=10) shows the maximum recall setting (99,60) with high workload. Thus, maximizing recall for selecting the ranking of top-$k$ alerts may not always lead to the low workload recommendation. It is a multi-objective optimization challenge. 

We design our optimization solution using \textit{Pareto Optimality} principle~\cite{ross1973economic}, given the lack of ground-truth data and knowledge during the time-critical times about the domain user preferences, which are often required to reach the best solution for multi-objective problems. Our two competing objectives are to achieve low workload and high recall (or low error rate) as illustrated above. An optimal solution would be Pareto optimal when it is not feasible to improve an objective without a penalty to another -- a \textit{non-dominating} solution. 
Formally, a vector of feasible decision variables $\hat x^{*}$  
is Pareto optimal if there does not exist another feasible decision vector $\hat x$ 
such that $f(\hat x) \le f(\hat x^{*})$ 
and $f'(\hat x) < f'(\hat x^{*})$ for at least one $f'$. 

\section{Experiments and Analysis}
\label{sec:experiments}

For a robust validation of our approach, 
we experiment with two 
schemes to generate $RW$ matrices: 
\begin{itemize}
\item \textit{Periodic} algorithm processes messages posted in the time window of past $H$ hours for generating top-$k$ ranking and a $RW$ matrix at the beginning of every hour (e.g., 7am, 8am). We consider $H=24$ for a fair recommendation of recall and the required workload from a generated $RW$. 
\item \textit{Realtime} algorithm processes messages posted in the time window of past $G$ minutes for generating top-$k$ ranking and a $RW$ matrix at the beginning of every minute (e.g., 7:01am, 7:02am). We consider $G=60$ for as accurate estimation as possible for the  observed 
$RW$.   
\end{itemize} 

\begin{figure*}
 \centering
  \subfloat[]{\includegraphics[width = 3.5in]{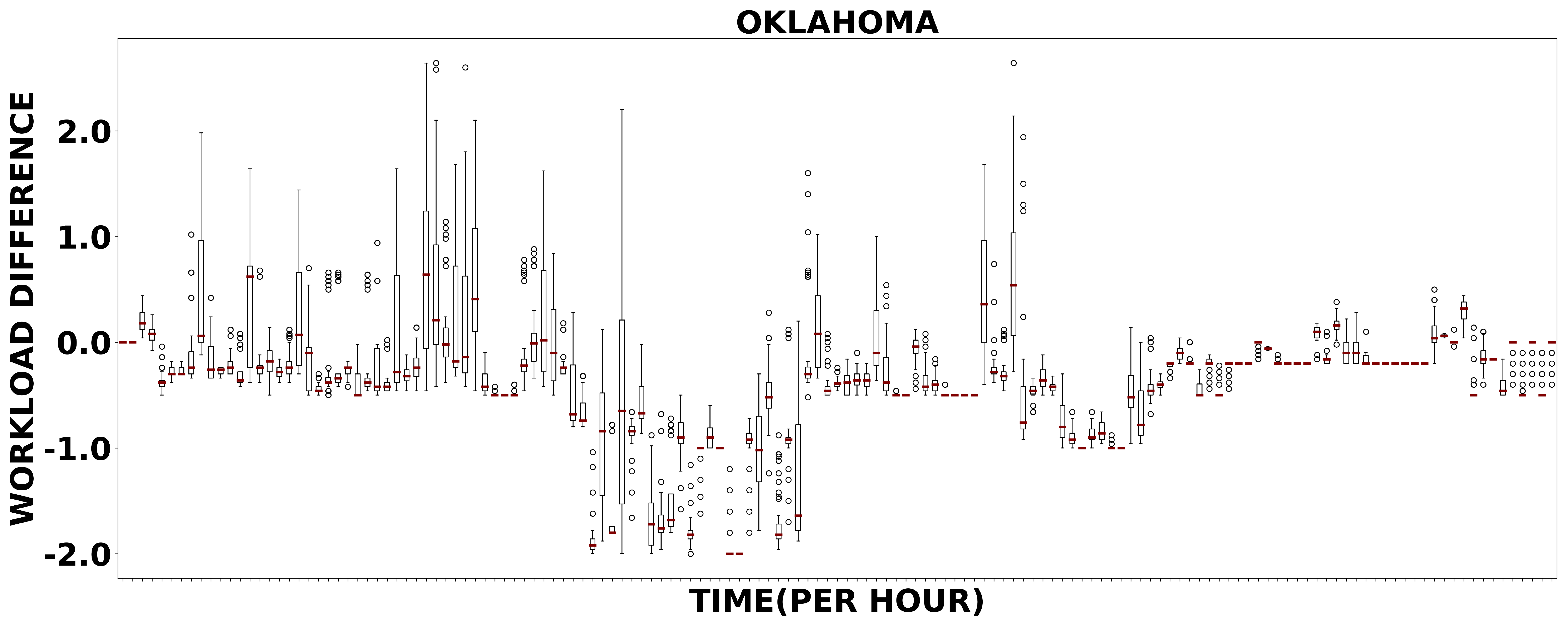}} 
  \subfloat[]{\includegraphics[width = 3.5in]{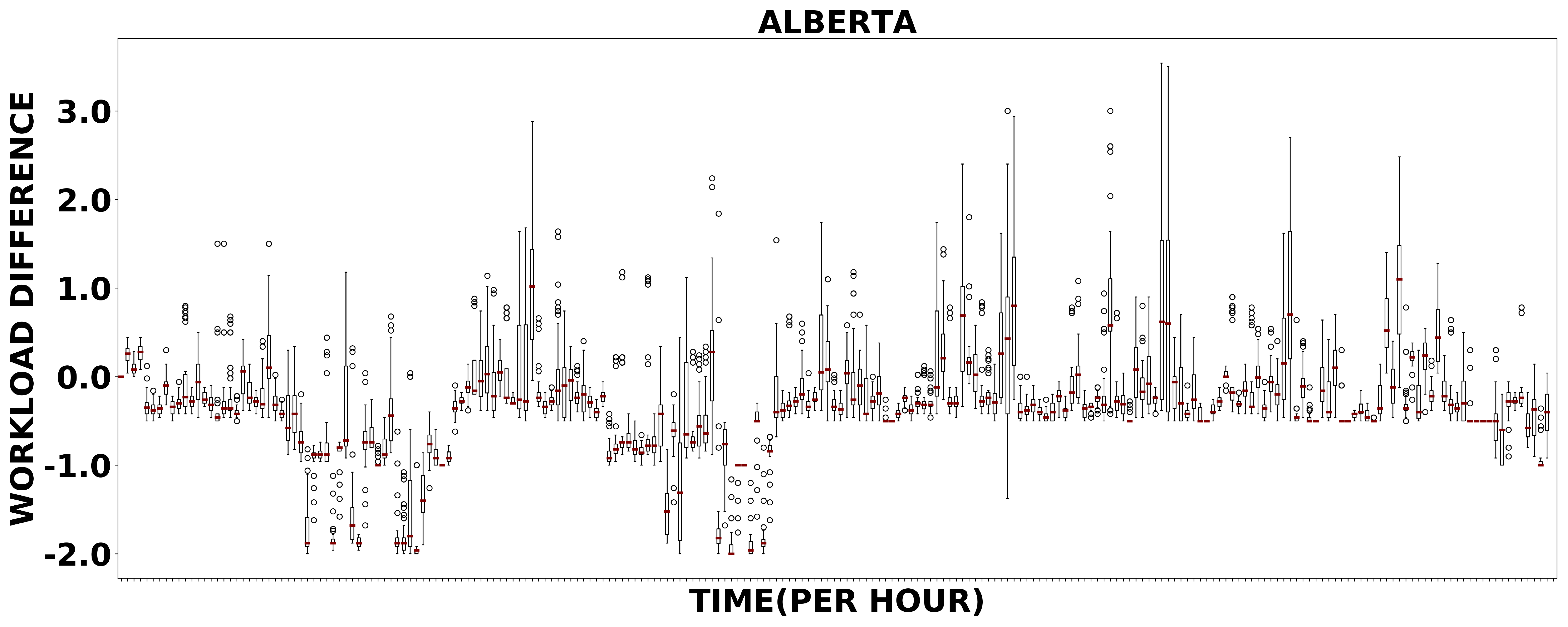}} \\
 \subfloat[]{\includegraphics[width = 3.5in]{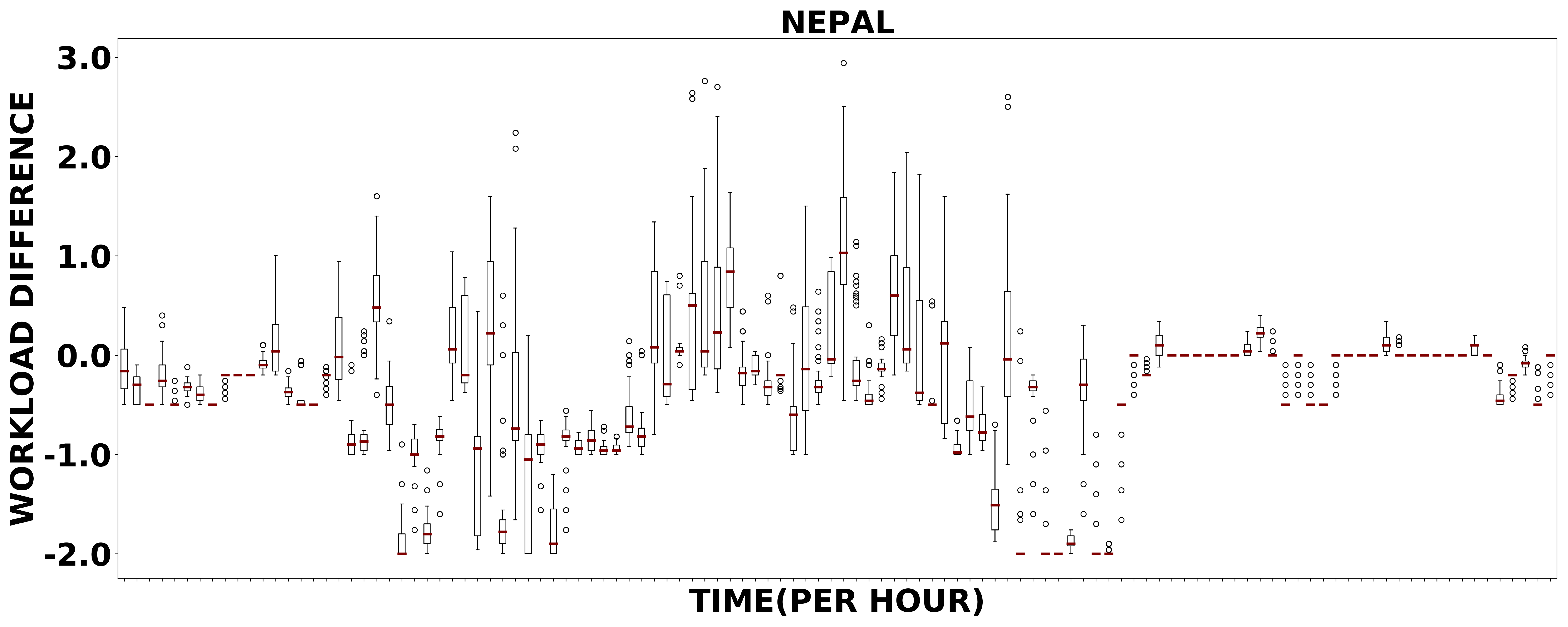}} 
 \subfloat[]{\includegraphics[width = 3.5in]{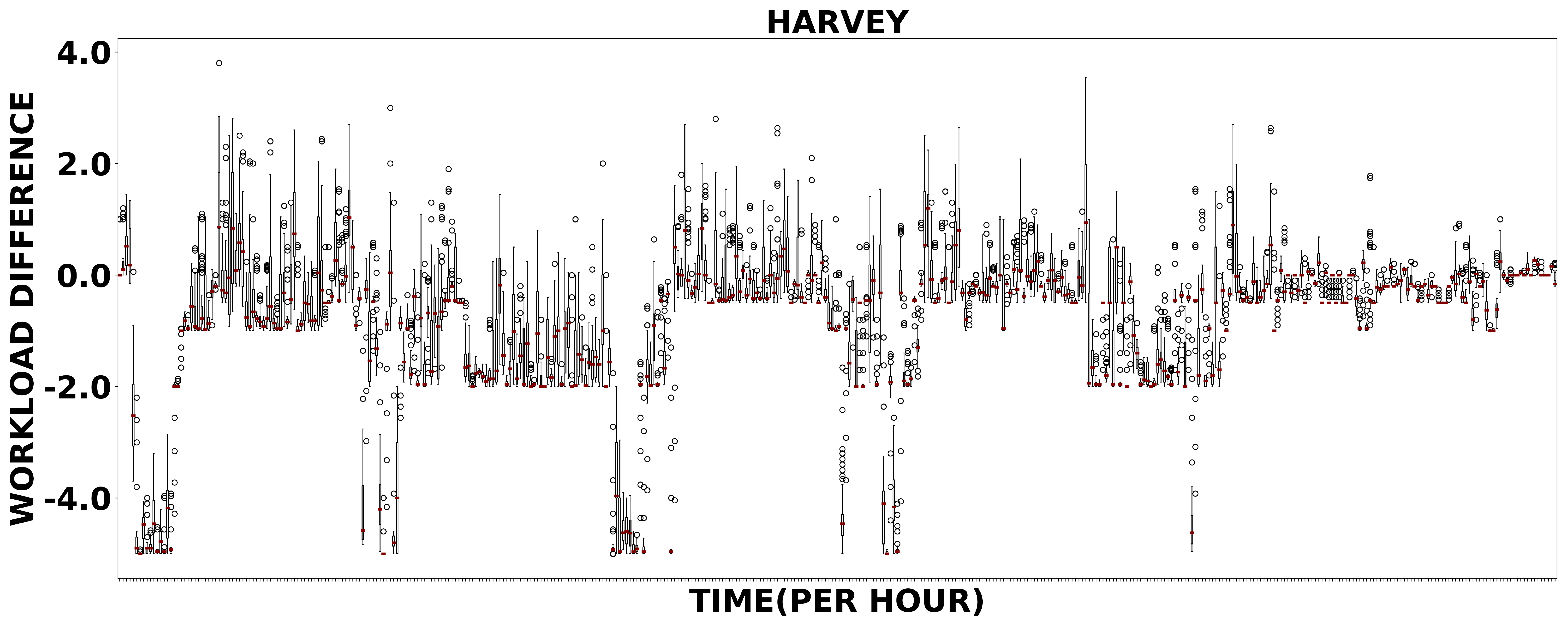}} 
 
 \caption{
 Moving average difference between the minimum workload recommendations obtained by the periodic and realtime algorithms shows very small error, indicating the effectiveness of our periodic approach to recommend workload. [\textit{Due to space limitation we skipped the figures of two smallest events.}]}
\label{fig:workload-boxplot}
\vskip -0.1in
\end{figure*}

We borrowed the datasets of 6 crisis events (\textit{c.f.} table~\ref{tab:dataset}) from our prior work~\cite{purohit2018socialeoc}. We processed them using the approach described in section~\ref{sec:approach}, within the periodic and realtime schemes. We used an existing epsilon-non-dominated sorting algorithm for the Pareto optimization~\cite{deb2005evaluating}. We analyze the following patterns for the relationship between the ranking performance and user workload achieved by the periodic and realtime schemes:  
patterns of recall versus workload recommendation and 
adaptive workload recommendation, and 
also, Pareto optimization comparison against the greedy baselines.    
\vskip -0.1in
\subsection{Recall vs Workload Recommendation Analysis} 
We studied the behavior of the average recall values for a value of workload and vice versa. In the highly stressful environment, the end users may not guarantee their availability for monitoring alerts consistently every hour. Thus, to support their decision making given such dynamic availability, the recommended $RW$ matrix provides how much  workload is necessary to achieve the desired recall (system performance) from the top-$k$ alert ranking. 
Based on the periodic scheme, we computed the average values of recall and workload obtained across all the time slices of an event as shown in figure~\ref{fig:workload-chull}.  

Event-specific sub-figures demonstrate that there exists a pattern of multiple recall values (corresponding to different top-$k$ rankings) for a given workload on x-axis, for instance, workload=10. The figures indicate the pattern of diminishing returns. 
The results also demonstrates the variability in the recall for the same workload across different events. Thus, one policy of selecting a specific top-$k$ ranking and desired recall cannot be applied to all the events consistently.  

\subsection{Adaptive Workload Recommendation Errors} 
The analytical goal here is 
to analyze the error 
between the periodic predictive recommendations and the near realtime values of the required workload. We also assess the Pareto optimization performance against the baseline ranking selections.   

\subsubsection{Error Analysis}
We first computed workload and recall for the hourly periodic algorithm output and then, the realtime algorithm output at every minute. We then measured the error difference between the outputs, followed by estimating the hourly mean and variance of error values. We observed that the error pattern was not contiguous for all the time slices across all the events.  
Therefore, we plotted the difference between the moving average values of each of the recall and workload metrics, where the average was computed across the sliding window of next 5 time periods. 
Figure~\ref{fig:workload-boxplot} 
shows the pattern of a stable moving average for the error ranges across all the events, implying that the proposed periodic algorithm tends to rectify the estimation error in the  recommendation in the near future. 
We further observed the bounded error ranges between the estimated and real values. The ranges are within 10\% of the maximum possible workload (60), thus, showing the potential of the proposed approach to recommend the optimal values of workload and recalls for top-$k$ alert rankings.   

\subsubsection{Baseline Comparison -- Greedy Selection} 
We analyze the difference between the performance metrics obtained by our Pareto approach and two biased, greedy baselines. 

\begin{figure*}
 \centering
  \subfloat[]{\includegraphics[trim={0 1.6cm 0 0}, width =  3.2in]{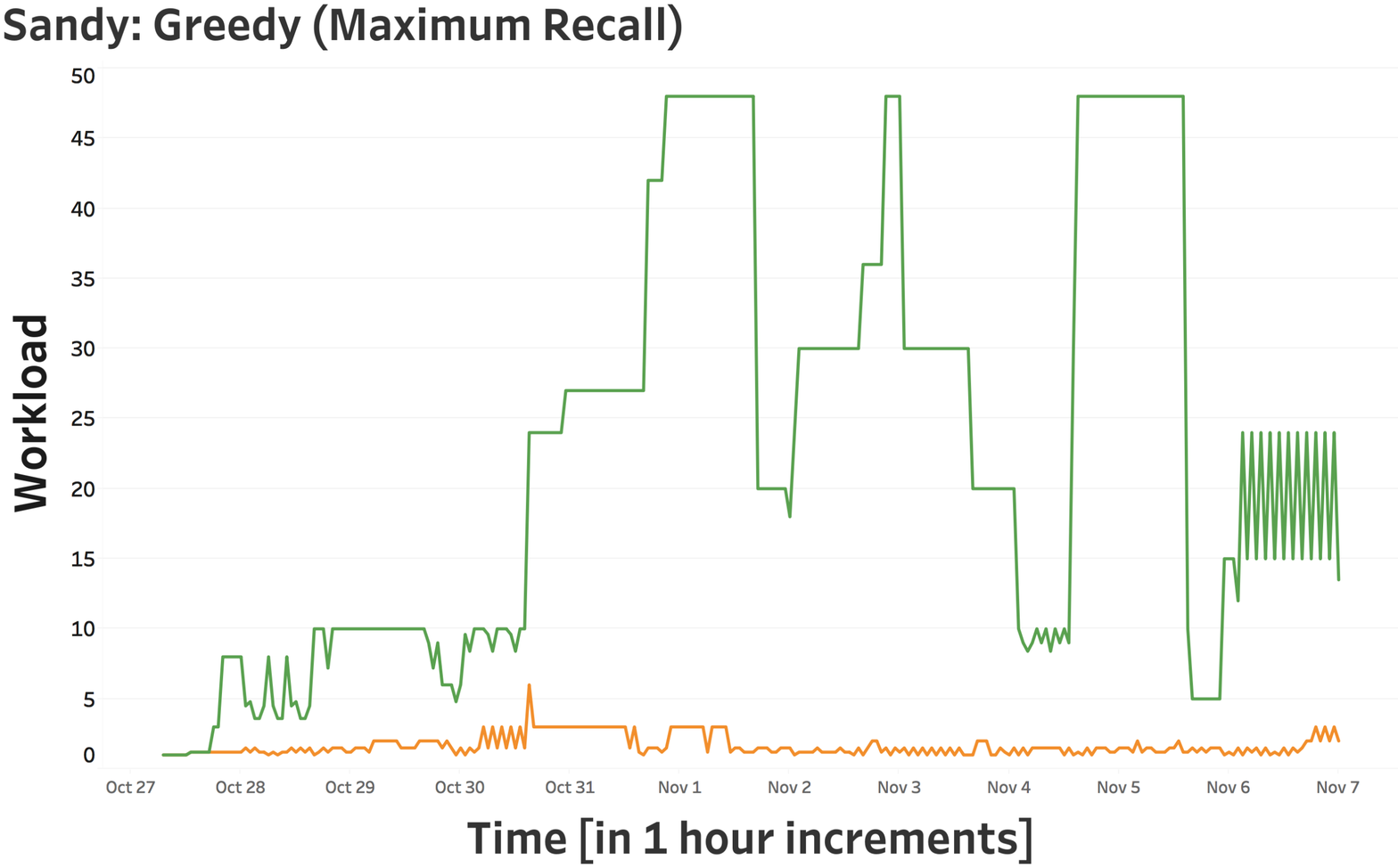}} 
 \subfloat[]{\includegraphics[trim={0 1.6cm 0 0}, width =  3.2in]{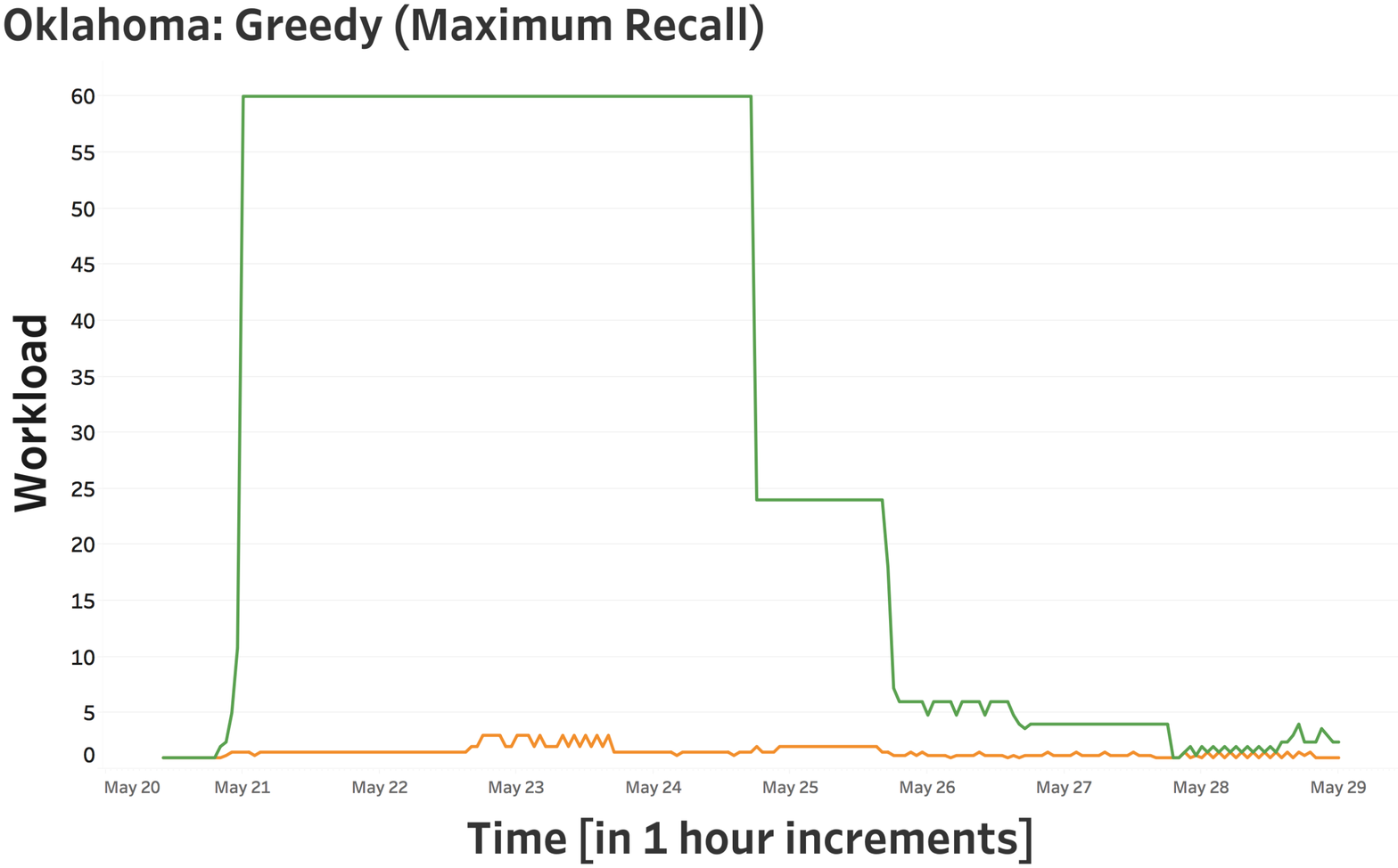}} \\
 \subfloat[]{\includegraphics[trim={0 1.6cm 0 0}, width =  3.2in]{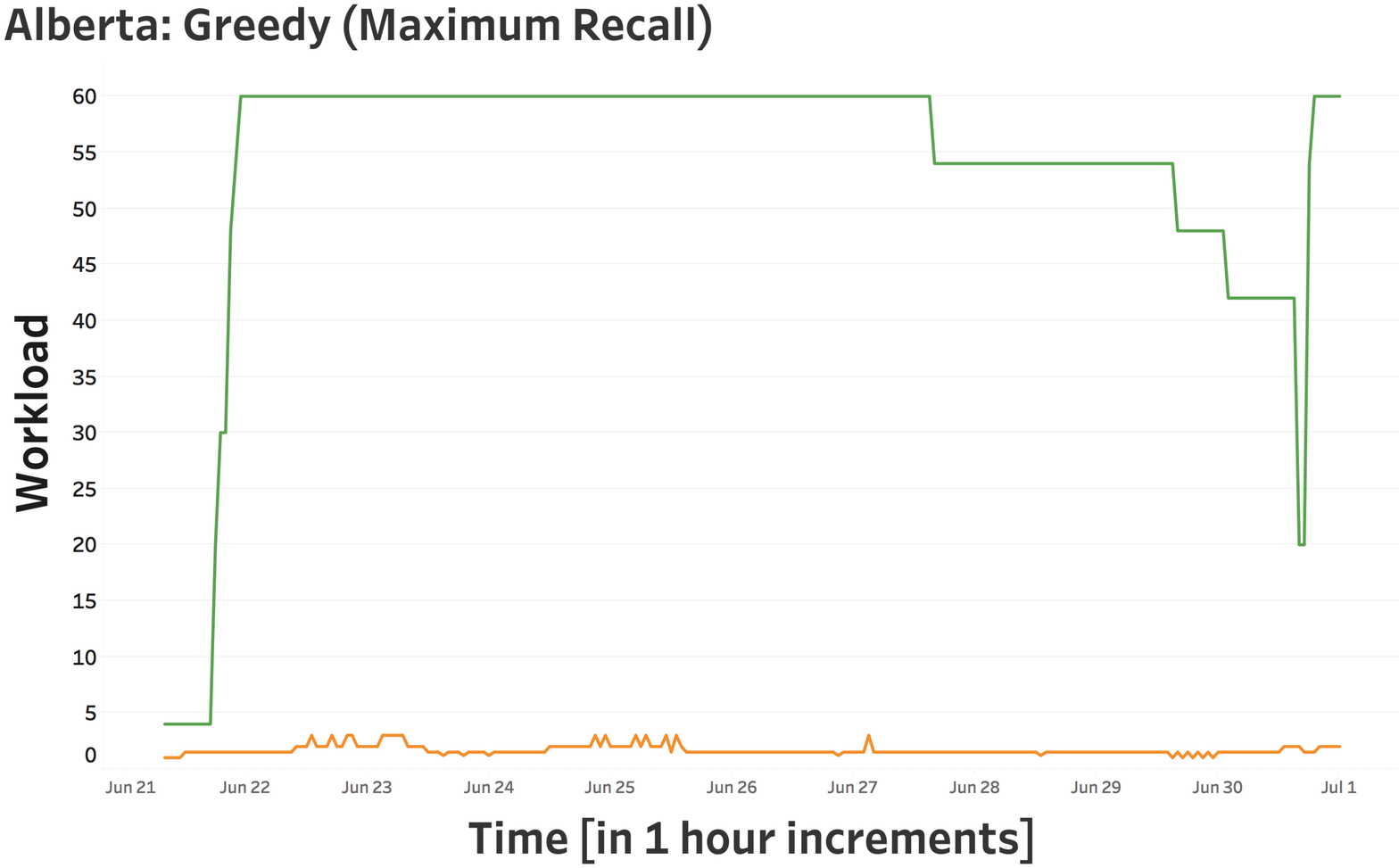}} 
 \subfloat[]{\includegraphics[trim={0 1.6cm 0 0}, width =  3.2in]{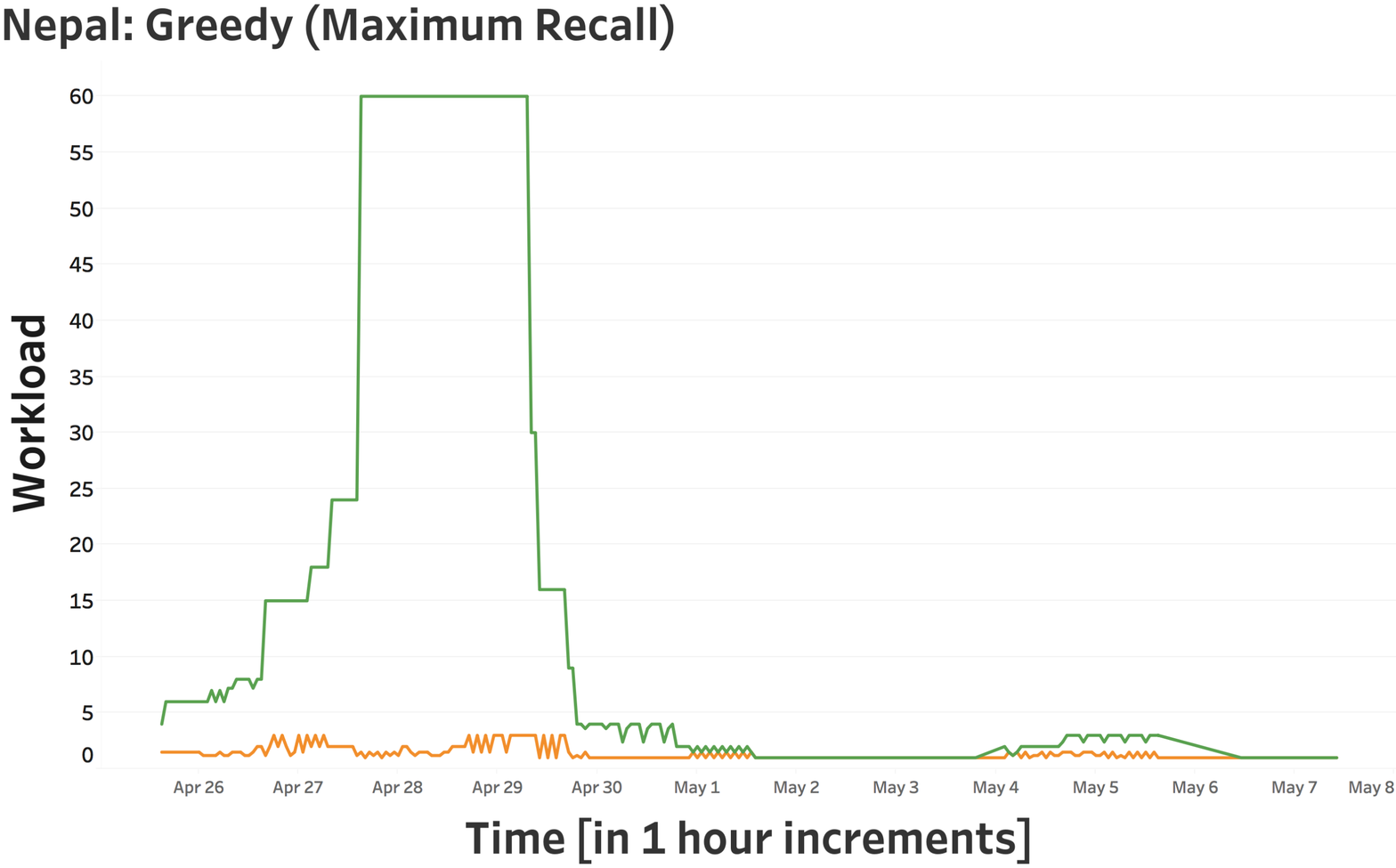}} \\
  \subfloat[]{\includegraphics[trim={0 1.6cm 0 0}, width =  3.2in]{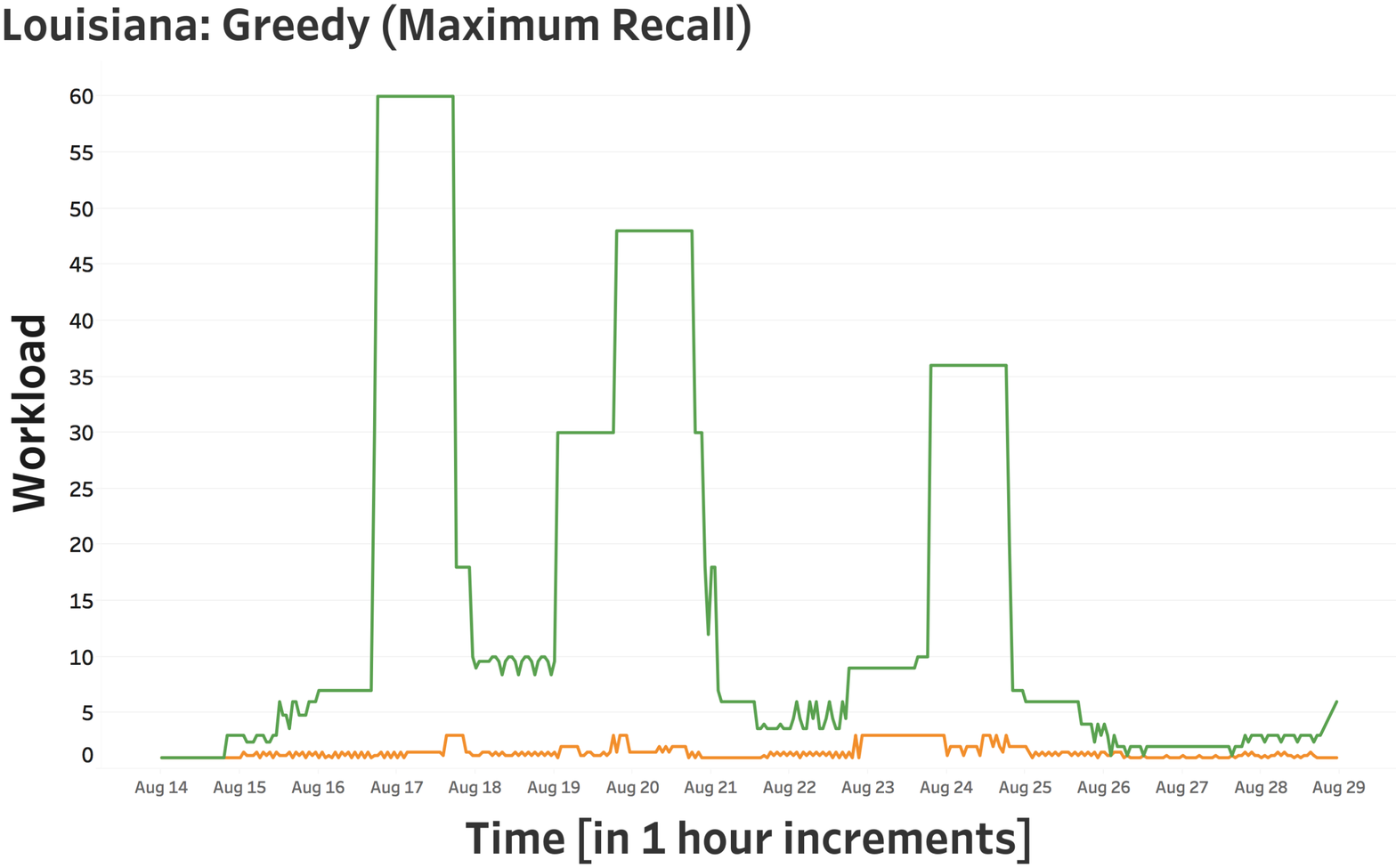}} 
 \subfloat[]{\includegraphics[trim={0 1.6cm 0 0}, width =  3.2in]{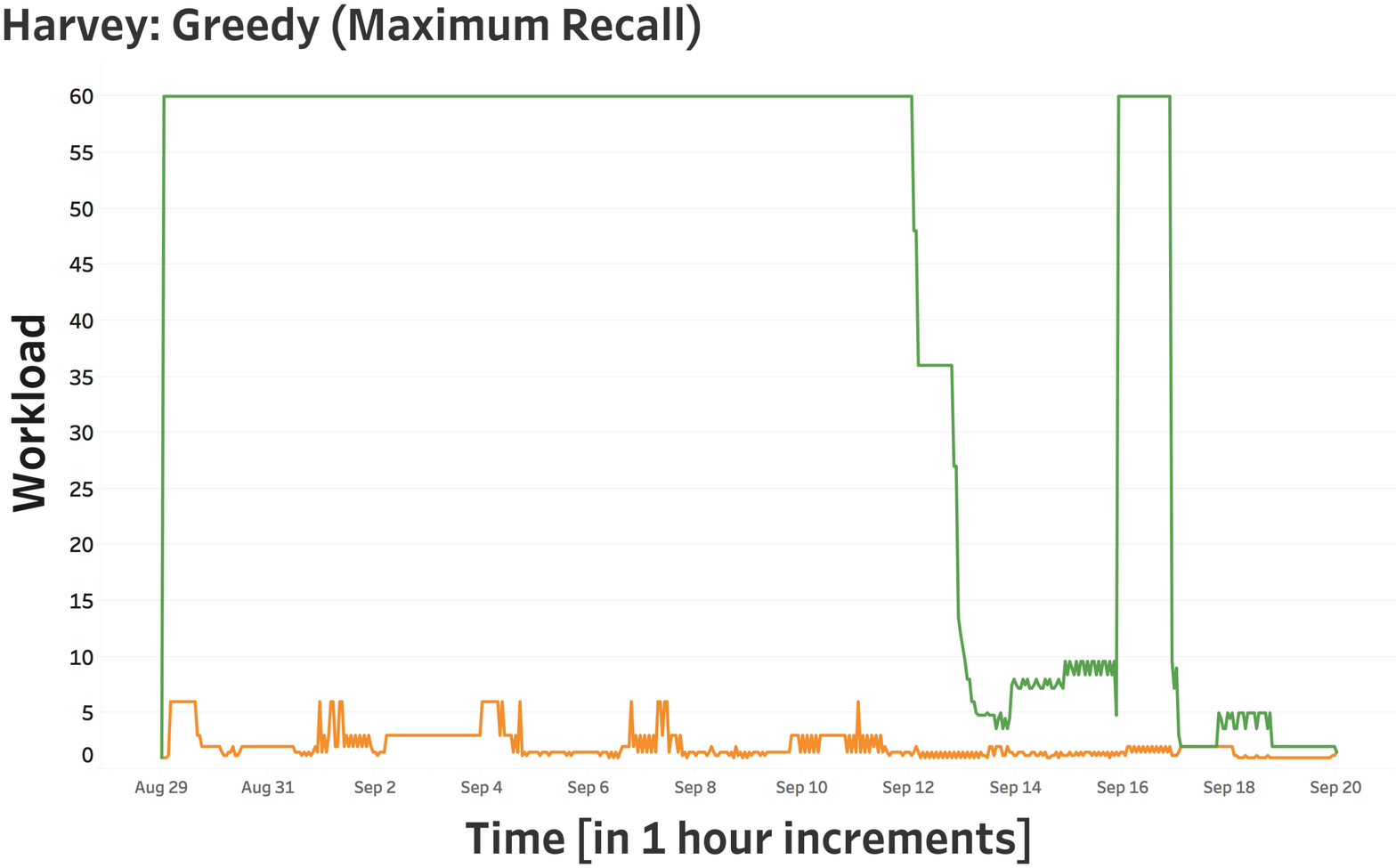}} 
 \caption{Lower value of minimum workload is recommended by the proposed Pareto algorithm (\textit{orange}) in comparison to that obtained by the baseline, recall-based Greedy algorithm (\textit{green}). } 
\label{fig:pareto-greedy-workload}
\end{figure*}

\begin{figure*}[t]
 \centering
  \subfloat[]{\includegraphics[trim={0 1.6cm 0 0}, width =  3in]{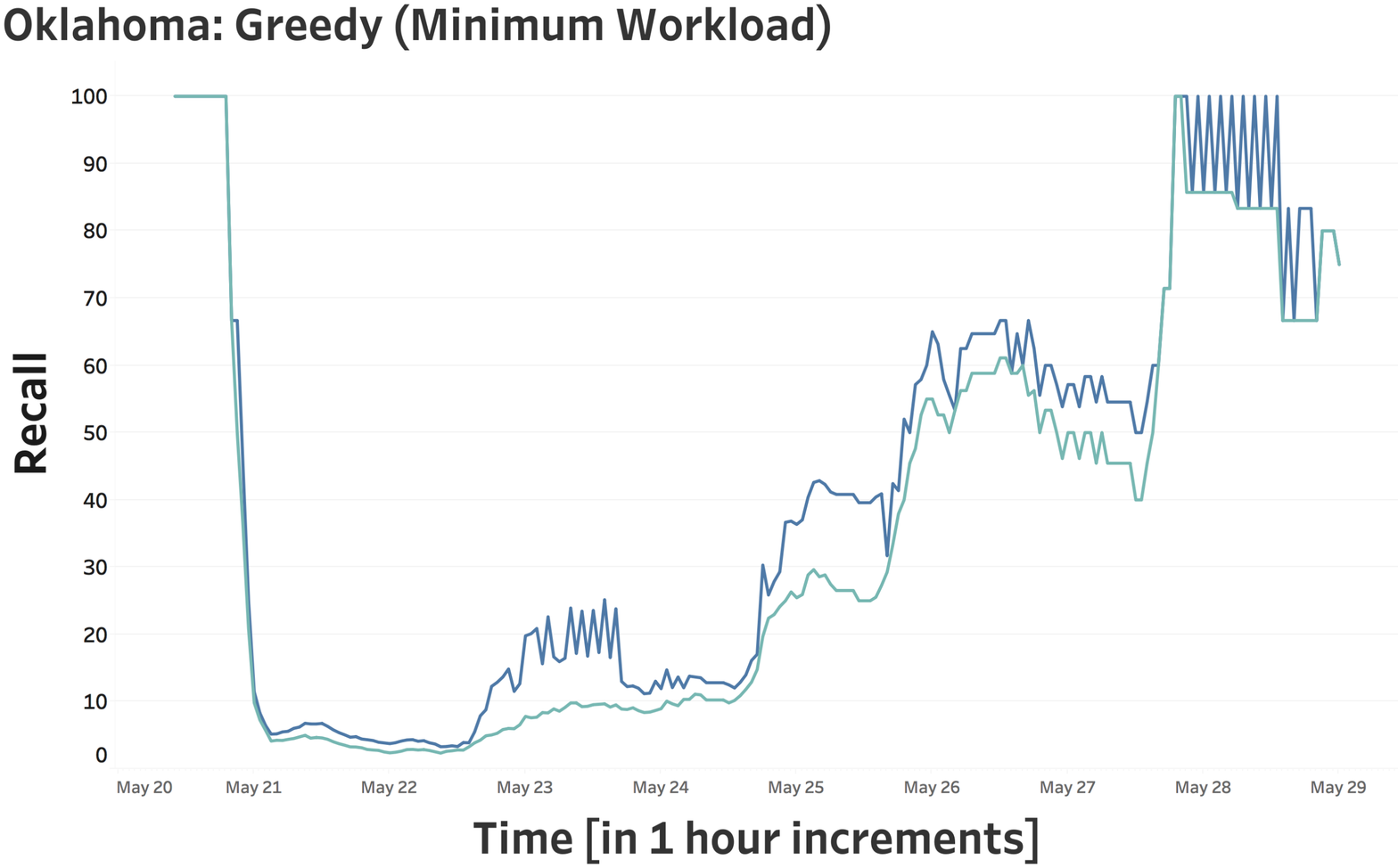}} 
  \subfloat[]{\includegraphics[trim={0 1.6cm 0 0}, width =  3in]{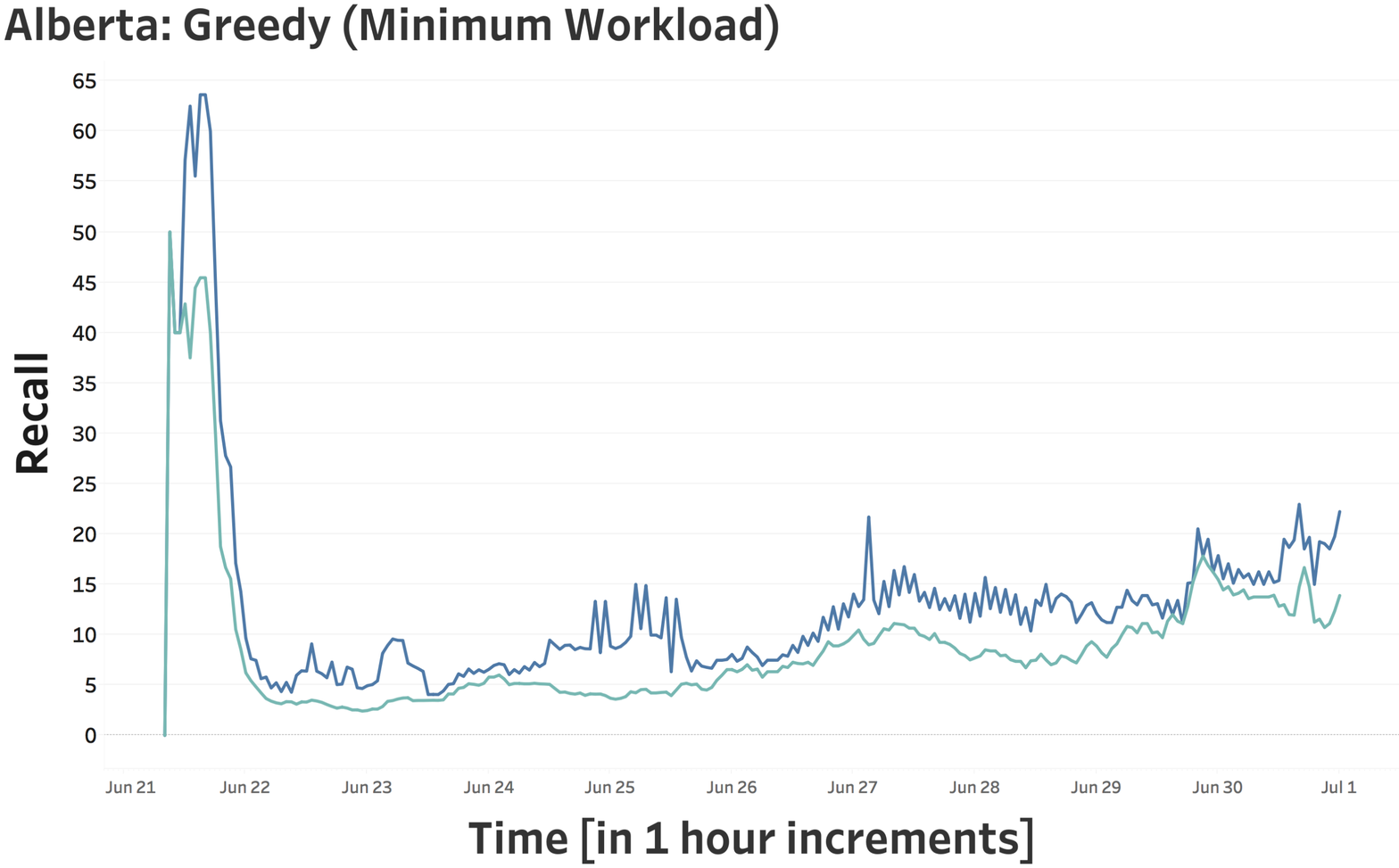}} \\
 \subfloat[]{\includegraphics[trim={0 1.6cm 0 0}, width =  3in]{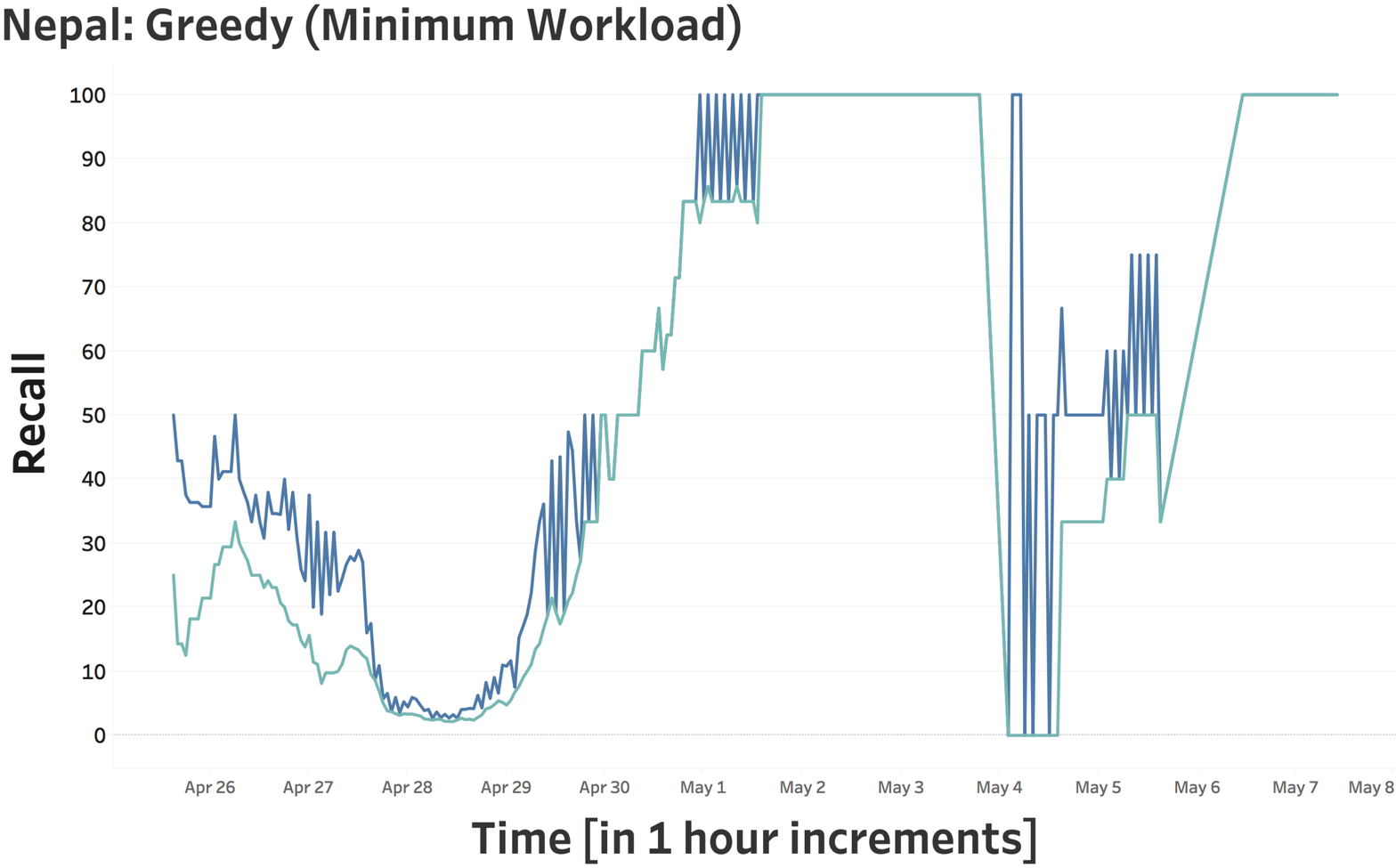}} 
 \subfloat[]{\includegraphics[trim={0 1.6cm 0 0}, width =  3in]{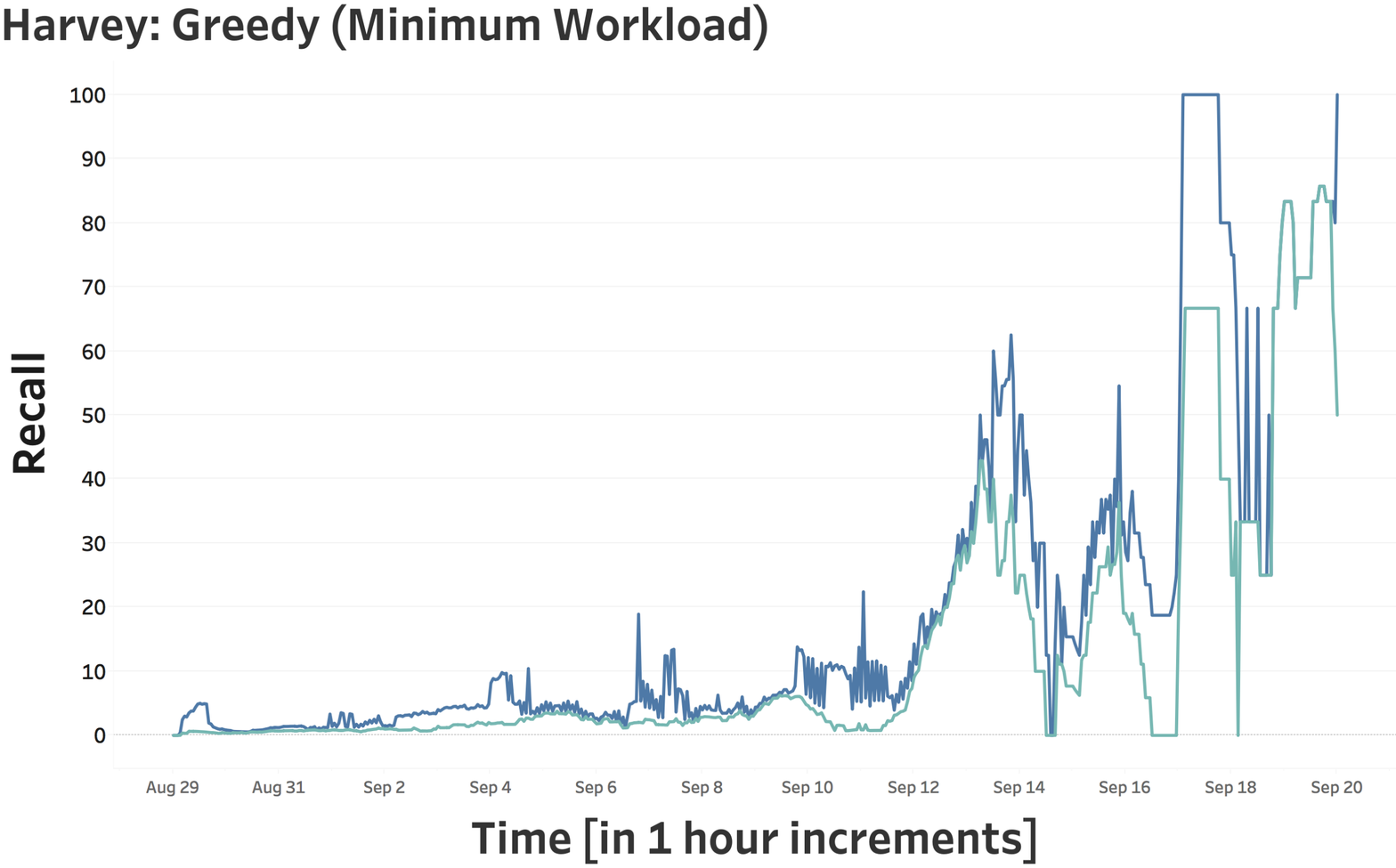}} 
 \caption{Higher value of maximum recall at the minimum workload is attainable by the proposed Pareto algorithm (\textit{blue}) in contrast to that obtained by the baseline, workload-based Greedy algorithm (\textit{green}). [\textit{We skipped smallest two events due to space limitation.}]} 
\label{fig:pareto-greedy-recall}
\end{figure*}

First greedy approach relies on the policy of selecting the alert ranking with minimum workload recommendation every time and the second one relies on selecting  the ranking with maximum recall recommendation. Figures~\ref{fig:pareto-greedy-workload} and~\ref{fig:pareto-greedy-recall} show the shortcoming of the greedy approaches where the choice of minimum feasible workload for recommendation does not always yield the maximum recall and thus, waste the time of the EOC personnel to review the irrelevant, useless messages.  

\begin{figure}[H]
\centering
\includegraphics[width=2.5in]{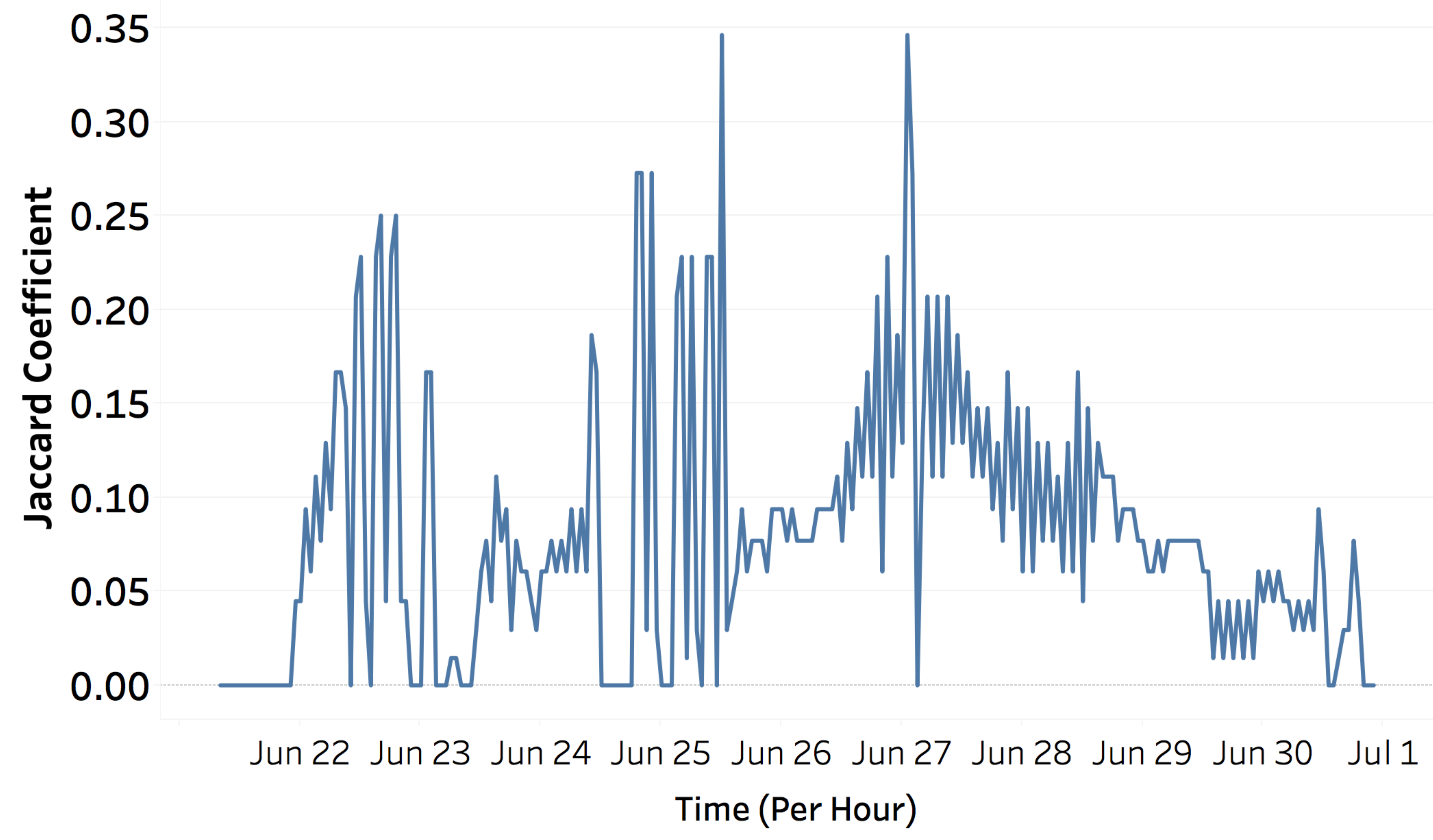} 
\caption{Illustration of redundancy 
by the overlap of current top-$50$ alerts in the top-$10$ ranked alerts of the two succeeding time periods for Alberta event.} 
\label{fig:alberta-redundancy} 
\vskip -0.2in
\end{figure}

\subsection{Redundancy and Timeliness} 
We also explored the information quality issues of the top-$k$ ranked messages in a time period $t_{ij}$ for redundancy. 
We computed redundancy using Jaccard Similarity between the set of top-$10$ alerts in $t^{s+1}_{ij}$ and $t^{s+2}_{ij}$ and the set of top-$50$ alerts in $t^{s}_{ij}$. Figure~\ref{fig:alberta-redundancy} shows the performance of periodic algorithm that re-surfaced the important messages as redundant alerts in future. 
This pattern suggests the need to efficiently factor redundancy and timeliness in the ranking computation. 

We also noticed that there are multiple choices in terms of which two points of $k$ (for top-$k$) and $t_{ij}$ (for period of computation) should be chosen. Table IV 
shows an illustration, if two points ($k$,$t_{ij}$) have the same workload, then we can create role-based, user-specific preference scheme. It is because every time the user receives an alert ranking list to review, s/he must switch work context. Therefore, we prefer to have lower frequency of requiring attention for user's review and recommend the 
smaller $k$ and larger $t_{ij}$ (less frequency to review without sacrificing performance).  

\section{Limitations}
\label{sec:discuss}
\begin{table*} 
  \label{tab:optimal-choices} 
  \centering
  \small
 \caption{Illustration of multiple choices of $k$ and $t_{ij}$ recommended for Sandy event by different ranking selection policies, which allow user-specific preference scheme. For instance, consider the desired recall of 60\% (\textit{bold row}), a user can choose smaller number of $k$ alerts and less frequency of $t_{ij}$ by the Pareto approach without sacrificing performance.} 
  \begin{tabular}{C{2cm}|C{2cm}|C{1.5cm}|C{1.5cm}|C{2cm}|C{2cm}|C{2cm}|C{2cm}} 
    \hline 
    Desired Recall &
    Recommended \newline Workload &
    Pareto $k$ &
    Pareto $t_{ij}$ &
    Greedy-Recall $k$ (baseline)&
    Greedy-Recall $t_{ij}$ (baseline)&
    Greedy- Workload $k$ (baseline) &
    Greedy- Workload $t_{ij}$ (baseline)\\ 
    \hline 
10\%    & 1.5  & 1 & 40  & 1 & 60  & 10 & 10 \\
20\% & 3    & 1 & 20  & 5 & 60  & 10 & 10 \\
.. & ..    & .. & ..  & .. & ..  & ..  & .. \\
40\% &  1    & 4 & 60  & 4 & 60  & 8  & 20 \\
\textbf{60}\% & \textbf{1}    & \textbf{1} & \textbf{60}  & \textbf{3} & \textbf{60}  & \textbf{4}  & \textbf{60} \\
.. & ..    & .. & ..  & .. & ..  & ..  & .. \\
  \hline 
\end{tabular}
\vskip -0.1in
\end{table*}

To the best of our knowledge this is a first study on the relationship between the performance metrics of alert ranking systems and the expected workload on end users in time-critical workplaces. Although this research  serves as a preliminary work towards future research on user-aware adaptive ranking methods. Given the scope of our study, the additional analyses can be addressed as future work. First, we did not explore different types of ranking systems for our analysis, it is possible that alert-based and static IR-based ranking systems would perform differently. Second, we demonstrated our analyses using a specific domain's data, i.e. emergency management, however, in other domains, the performance might vary in terms of the error between the estimated and real recommendations. Third, we have not studied the large range of workload bounds (considered only a reasonable range from 1 to 60 messages/hour) and its effects on the communications officers in EOC (e.g., one could hypothesize an excessive cognitive workload in the upper part of this range), which could be further studied. 
Lastly, the presented results depend on the relevance-based ranking system, however, there is a possibility to also incorporate redundancy and timeliness factors in the ranking system, which can be explored as a future work.    
\vskip -0.2in

\section{Conclusions}
\label{sec:conclusion}
Due to the limited budget of attention in the stressful environment of emergency management, traditional `one-size-fits-all' solutions 
of alert generation for relevant social media updates are not effective. This paper presented a novel quantitative model for determining how many and how often should social media updates be generated, while also considering a given bound on the workload for an end user. 
Our formal model quantifies the relationship between the performance metrics of recall for top-$k$ rankings and the required user workload. We presented an alert ranking system that employs a Pareto optimal algorithm for ranking selection, by adaptively determining the preference of top-\textit{k} ranking and user workload over time.     
We presented empirical results based on real-world data from 6 crisis events to study the effects of different ranking selections and the trade-off with user workload, in comparison to different greedy baseline approaches.   
Our experiments demonstrate that the proposed approach  
can improve the efficiency of monitoring social media updates for EOC personnel while respecting constraints in user attention.  

\spara{Reproducibility.} \textit{Our dataset is available upon request, for research purposes.} 
\section{Acknowledgement}
Authors would like to thank reviewers for valuable feedback. Also, Purohit thanks US National Science Foundation grants IIS-1657379 \& IIS-1815459 and Castillo thanks La Caixa project LCF/PR/PR16/11110009 for partial support. 

\bibliographystyle{IEEEtran}
\bibliography{paper-wi-workload} 

\begin{thebibliography}{10}
\providecommand{\url}[1]{#1}
\csname url@samestyle\endcsname
\providecommand{\newblock}{\relax}
\providecommand{\bibinfo}[2]{#2}
\providecommand{\BIBentrySTDinterwordspacing}{\spaceskip=0pt\relax}
\providecommand{\BIBentryALTinterwordstretchfactor}{4}
\providecommand{\BIBentryALTinterwordspacing}{\spaceskip=\fontdimen2\font plus
\BIBentryALTinterwordstretchfactor\fontdimen3\font minus
  \fontdimen4\font\relax}
\providecommand{\BIBforeignlanguage}[2]{{%
\expandafter\ifx\csname l@#1\endcsname\relax
\typeout{** WARNING: IEEEtran.bst: No hyphenation pattern has been}%
\typeout{** loaded for the language `#1'. Using the pattern for}%
\typeout{** the default language instead.}%
\else
\language=\csname l@#1\endcsname
\fi
#2}}
\providecommand{\BIBdecl}{\relax}
\BIBdecl

\bibitem{arc_2012_survey}
{American Red Cross}, ``More americans using mobile apps in emergencies,''
  August 2012, online and phone survey.

\bibitem{hughes2012evolving}
A.~L. Hughes and L.~Palen, ``The evolving role of the public information
  officer: An examination of social media in emergency management,''
  \emph{Journal of Homeland Security and Emergency Management}, vol.~9, no.~1,
  2012.

\bibitem{castillo2016big}
C.~Castillo, \emph{Big Crisis Data: Social Media in Disasters and Time-Critical
  Situations}.\hskip 1em plus 0.5em minus 0.4em\relax Cambridge University
  Press, 2016.

\bibitem{purohit2013emergency}
H.~Purohit, C.~Castillo, F.~Diaz, A.~Sheth, and P.~Meier, ``Emergency-relief
  coordination on social media: Automatically matching resource requests and
  offers,'' \emph{First Monday}, vol.~19, no.~1, 2013.

\bibitem{he2017signal}
X.~He, D.~Lu, D.~Margolin, M.~Wang, S.~E. Idrissi, and Y.-R. Lin, ``The signals
  and noise: Actionable information in improvised social media channels during
  a disaster,'' in \emph{WebSci}, 2017, pp. 33--42.

\bibitem{starbird2014rumors}
K.~Starbird, J.~Maddock, M.~Orand, P.~Achterman, and R.~M. Mason, ``Rumors,
  false flags, and digital vigilantes: Misinformation on twitter after the 2013
  boston marathon bombing,'' \emph{iConference}, 2014.

\bibitem{purohit2018socialeoc}
\BIBentryALTinterwordspacing
H.~Purohit, C.~Castillo, M.~Imran, and R.~Pandey, ``Social-eoc: Serviceability
  model to rank social media requests for emergency operation centers,'' in
  \emph{ASONAM}, 2018, to appear. [Online]. Available:
  \url{http://ist.gmu.edu/~hpurohit/informatics-lab/papers/serviceability_ranking_disasters_ASONAM18_final.pdf}
\BIBentrySTDinterwordspacing

\bibitem{imran2015processing}
M.~Imran, C.~Castillo, F.~Diaz, and S.~Vieweg, ``Processing social media
  messages in mass emergency: A survey,'' \emph{ACM Computing Surveys},
  vol.~47, no.~4, p.~67, 2015.

\bibitem{avvenuti2014ears}
M.~Avvenuti, S.~Cresci, A.~Marchetti, C.~Meletti, and M.~Tesconi, ``Ears
  (earthquake alert and report system): a real time decision support system for
  earthquake crisis management,'' in \emph{KDD}, 2014, pp. 1749--1758.

\bibitem{aslam2015trec}
J.~Aslam, F.~Diaz, M.~Ekstrand-Abueg, R.~McCreadie, V.~Pavlu, and T.~Sakai,
  ``Trec 2014 temporal summarization track overview,'' NIST, Tech. Rep., 2015.

\bibitem{hiltz2014use}
S.~R. Hiltz, J.~A. Kushma, and L.~Plotnick, ``Use of social media by us public
  sector emergency managers: Barriers and wish lists.'' in \emph{ISCRAM}, 2014,
  pp. 602--611.

\bibitem{plotnick2015red}
L.~Plotnick, S.~R. Hiltz, J.~A. Kushma, and A.~H. Tapia, ``Red tape: Attitudes
  and issues related to use of social media by us county-level emergency
  managers.'' in \emph{ISCRAM}, 2015, pp. 182--192.

\bibitem{baeza2011ribeiro}
R.~Baeza-Yates and B.~Ribeiro-Neto, \emph{Modern Information Retrieval: The
  Concepts and Technology Behind Search}, 2nd~ed.\hskip 1em plus 0.5em minus
  0.4em\relax USA: Addison-Wesley Publishing Company, 2008.

\bibitem{rudin2009p}
C.~Rudin, ``The p-norm push: A simple convex ranking algorithm that
  concentrates at the top of the list,'' \emph{Journal of Machine Learning
  Research}, vol.~10, no. Oct, pp. 2233--2271, 2009.

\bibitem{palen2016crisis}
L.~Palen and K.~M. Anderson, ``Crisis informatics—new data for extraordinary
  times,'' \emph{Science}, vol. 353, no. 6296, pp. 224--225, 2016.

\bibitem{dhs2014using}
S.~M. W.~G. U.S. Homeland Security Science \&~Technology, ``Using social media
  for enhanced situational awareness and decision support,''
  \url{https://www.dhs.gov/publication/using-social-media-enhanced-situational-awareness-decision-support},
  2014, accessed: 2018-06-12.

\bibitem{reuter2017towards}
C.~Reuter and T.~Spielhofer, ``Towards social resilience: A quantitative and
  qualitative survey on citizens' perception of social media in emergencies in
  europe,'' \emph{Technological Forecasting and Social Change}, vol. 121, pp.
  168--180, 2017.

\bibitem{sakaki2013tweet}
T.~Sakaki, M.~Okazaki, and Y.~Matsuo, ``Tweet analysis for real-time event
  detection and earthquake reporting system development,'' \emph{IEEE TKDE},
  vol.~25, no.~4, pp. 919--931, 2013.

\bibitem{earle2012twitter}
P.~S. Earle, D.~C. Bowden, and M.~Guy, ``Twitter earthquake detection:
  earthquake monitoring in a social world,'' \emph{Annals of Geophysics},
  vol.~54, no.~6, 2012.

\bibitem{robinson2013sensitive}
B.~Robinson, R.~Power, and M.~Cameron, ``A sensitive twitter earthquake
  detector,'' in \emph{WWW}, 2013, pp. 999--1002.

\bibitem{yin2012using}
J.~Yin, A.~Lampert, M.~Cameron, B.~Robinson, and R.~Power, ``Using social media
  to enhance emergency situation awareness,'' \emph{IEEE Intelligent Systems},
  vol.~27, no.~6, pp. 52--59, 2012.

\bibitem{bao2013georank}
J.~Bao and M.~F. Mokbel, ``Georank: an efficient location-aware news feed
  ranking system,'' in \emph{GIS}, 2013, pp. 184--193.

\bibitem{kedzie2015predicting}
C.~Kedzie, K.~McKeown, and F.~Diaz, ``Predicting salient updates for disaster
  summarization,'' in \emph{ACL}, vol.~1, 2015, pp. 1608--1617.

\bibitem{mccreadie2014incremental}
R.~McCreadie, C.~Macdonald, and I.~Ounis, ``Incremental update summarization:
  Adaptive sentence selection based on prevalence and novelty,'' in
  \emph{CIKM}, 2014, pp. 301--310.

\bibitem{rudra2015extracting}
K.~Rudra, S.~Ghosh, N.~Ganguly, P.~Goyal, and S.~Ghosh, ``Extracting
  situational information from microblogs during disaster events: a
  classification-summarization approach,'' in \emph{CIKM}, 2015, pp. 583--592.

\bibitem{nenkova2011automatic}
A.~Nenkova and K.~McKeown, ``Automatic summarization,'' \emph{Foundations and
  Trends{\textregistered} in Information Retrieval}, vol.~5, no. 2--3, 2011.

\bibitem{sachdeva2017call}
N.~Sachdeva and P.~Kumaraguru, ``Call for service: Characterizing and modeling
  police response to serviceable requests on facebook,'' in \emph{CSCW}, 2017,
  pp. 336--352.

\bibitem{liu2009learning}
T.-Y. Liu, ``Learning to rank for information retrieval,'' \emph{Foundations
  and Trends{\textregistered} in Information Retrieval}, vol.~3, no.~3, pp.
  225--331, 2009.

\bibitem{ross1973economic}
S.~A. Ross, ``The economic theory of agency: The principal's problem,''
  \emph{The American Economic Review}, vol.~63, no.~2, pp. 134--139, 1973.

\bibitem{deb2005evaluating}
K.~Deb, M.~Mohan, and S.~Mishra, ``Evaluating the $\varepsilon$-domination
  based multi-objective evolutionary algorithm for a quick computation of
  pareto-optimal solutions,'' \emph{Evolutionary computation}, vol.~13, no.~4,
  pp. 501--525, 2005.

\end{thebibliography}

\end{document}